\documentclass[aps,twocolumn,preprintnumbers,amsmath,amssymb,superscriptaddress]{revtex4}
\usepackage[utf8]{inputenc}
\usepackage{amsmath}
\usepackage{amsfonts}
\usepackage{amssymb}
\usepackage{bm}
\usepackage{textcomp}
\usepackage{graphicx}
\usepackage{units}
\usepackage{hyperref}
\hypersetup{
	pdftitle=Thermodynamic stability of mixed Pb:Sn methyl-ammonium halide perovskites,
	pdfauthor={Ksenia Korshunova, Lars Winterfeld, Wichard J.D. Beenken and Erich Runge},
	pdfkeywords={Perovskite phases, hybrid solar cells, ab-initio calculations, entropy of mixing}
}

\begin{document}

\title{Thermodynamic stability of mixed Pb:Sn methyl-ammonium halide perovskites}
\author{Ksenia Korshunova}
\affiliation{Institut für Physik and Institut f\"{u}r Mikro- und Nanotechnologie, Technische Universit\"{a}t Ilmenau, 98684 Ilmenau, Germany}
\author{Lars Winterfeld}
\affiliation{Institut für Physik and Institut f\"{u}r Mikro- und Nanotechnologie, Technische Universit\"{a}t Ilmenau, 98684 Ilmenau, Germany}
\author{Wichard J.D. Beenken}
\affiliation{Institut für Physik and Institut f\"{u}r Mikro- und Nanotechnologie, Technische Universit\"{a}t Ilmenau, 98684 Ilmenau, Germany}
\author{Erich Runge}
\affiliation{Institut für Physik and Institut f\"{u}r Mikro- und Nanotechnologie, Technische Universit\"{a}t Ilmenau, 98684 Ilmenau, Germany}

\date{June 20, 2016}

\newcommand{\MA}{M\hspace{-0.35ex}A\hspace{0.35ex}}
\newcommand{\FA}{F\hspace{-0.35ex}A\hspace{0.35ex}}
\hyphenation{cub-octahedral}

\begin{abstract}
Using density functional theory, we investigate systematically mixed $\MA(Pb:Sn)X_3$ perovskites, where $\MA$ is $CH_3NH_3^+$, and $X$ is $Cl$, $Br$ or $I$.
Ab initio calculations of the orthorhombic, tetragonal and cubic perovskite phases show that the substitution of lead by tin has a much weaker influence on both structure and cohesive energies than the substitution of the halogen.
The thermodynamic stability of the $\MA(Pb:Sn)X_3$ mixtures at finite, non-zero temperatures is studied within the Regular Solution Model.
We predict that it will be possible to create $\MA(Pb:Sn)I_3$ mixtures at any temperature.
Our results imply that mixing is unlikely for the low-temperature phase of bromide and chloride compounds, where instead local clusters are more likely to form.
We further predict that in the high-temperature cubic phase, $Pb$ and $Sn$ compounds will mix for both $\MA(Pb:Sn)Br_3$ and $\MA(Pb:Sn)Cl_3$ due to the entropy contribution to the Helmholtz free energy.
\end{abstract}

\maketitle
\onecolumngrid
\textit{This is the peer reviewed version of the following article: Phys. Status Solidi B, 253: 1907--1915, which has been published in final form at \href{http://dx.doi.org/10.1002/pssb.201600136}{DOI: 10.1002/pssb.201600136}. This article may be used for non-commercial purposes in accordance with Wiley Terms and Conditions for Self-Archiving.}\\
\twocolumngrid

\section{Introduction}

Solar cells based on hybrid organic-inorganic perovskites as light absorbers have recently received an enormous interest due to their potential to be manufactured much cheaper than monocrystalline silicon cells using, e.g., wet chemistry and spin coating \cite{Lee2012,Kim2012,Noh2013,DeBastiani2014}.
In this context, the term "hybrid organic-inorganic perovskites" refers to a broad class of metal halides that can be described with the general chemical formula $ABX_3$, where $A$ is an organic cation, such as (but not limited to) methylammonium ($CH_3NH_3^+$, MA)\cite{Baikie2013,Giorgi2013,Weller2015}, ethylammonium ($CH_3CH_2NH_3^+$, EA)\cite{Im2012,Safdari2014}, formamidinium ($HC(NH_2)_2^+$, FA)\cite{Borriello2008,Stoumpos2013b,Pellet2014} or guanidinium ($CH_6N_3^+$, GA)\cite{Giorgi2016,Marco2016}, $B$ is a metal cation (usually lead), and $X$ is a halide anion.
The metal cations are coordination centers of $BX_6$ octahedra.
We focus on compounds, which build a corner-connected three-dimensional network.
The organic cations $A$ are located in the cuboctahedral cavities between the $BX_6$ octahedra.
Therefore, their size is limited to fit the cavity without major distortions of the octahedral network.
Since the stability of the organic-inorganic superlattice strongly depends on the size of $A$, the relatively small MA-cation is one of the most commonly used organic compounds.
Lead-based perovskite solar cells with MA as organic cation reach electrical power conversion efficiencies (PCE) of 16\%\cite{Ryu2014,Jeon2014,Lee2014}, and in the most recently published reports the PCE reaches 18\% for mesoporous $MAPbI_3$.\cite{Bi2016}
In the case of $FAPbI_3$-based solar cells, the achieved maximum PCE is greater than 20\%.\cite{Yang2015} 
However, in view of the potential toxicity of lead, substitution of lead with another group 14 metal, e.g., tin, seems desirable.
The search for optimized photoelectric properties initiated a lot of experimental and theoretical research on mixed-halide structures,\cite{Noh2013,Mosconi2013,Hoke2014,DeBastiani2014,Zheng2015} or structures with mixed organic cations.\cite{Pellet2014,Liu2015}
The highest reported efficiencies at the moment are found for combinations of various organohalide perovskites, such as $FAPbI_3$ mixed with $MAPbBr_3$ with the impressive PCE of 20.1\%.\cite{Seo2016}
Much less is known about the effects of incomplete substitution of lead by other metals.
The reason is most likely that researchers either prefer to stick with the very successful ${\MA}PbI_3$\cite{Wang2014b} or wish to get rid of Pb altogether.\cite{Hao2014b,Noel2014,Bernal2014}
Only recently, the mixed ${\MA}Sn_xPb_{1-x}I_3$ was reported to have a spectrally extended absorption compared to the non-mixed lead-halide perovskites, shifting the absorption onset down to the near infrared.\cite{Hao2014,Zuo2014,Ogomi2014}
Furthermore, effective-mass tuning for improved transport properties was suggested for mixed \textit{Pb:Sn} iodide compounds in the cubic phase.\cite{Mosconi2014}
We think that mixtures with respect to the metal component deserves to be studied in more detail.
A better understanding of the whole class of hybrid organohalide perovskites may in the end even lead to Pb-free high-performance cells.

In this paper, we provide a systematic \textit{ab-initio} investigation of ${\MA}BX_3$ perovskites, where $B$ is $Sn^{2+}$, $Pb^{2+}$ or various mixtures thereof, and $X$ is $I^-$, $Br^-$ or $Cl^-$.
The starting point is the discussion of general structural and energetic properties of the non-mixed structures in three phases (orthorhombic, tetragonal, and cubic, known from ${\MA}PbI_3$ at different temperatures) depending on their chemical composition.
We then introduce a variety of mixed Sn-Pb structures, which differ not only by the total Sn:Pb-ratio but also by the arrangement of the metal atoms in the unit cell.
Based on the calculated cohesive energies of these mixed compounds, we employ the Regular Solution Model in order to investigate their thermodynamic stability.
Details are described in the next section.

\section{Methodology}

We used density functional theory (DFT) to calculate the energy for several atomic structures, as described in the first subsection below.
However, DFT alone does not account for entropy contributions, which are relevant for finite temperature predictions.
Therefore, a thermodynamic model is needed, which is described in the second subsection.

\subsection{Electronic structure and geometry optimization}

To optimize crystal structures and calculate binding energies, we performed \textit{ab initio} DFT calculations using the Vienna Ab initio Simulation Package (VASP).\cite{Kresse1993,Kresse1996}
Projector-augmented wave (PAW)\cite{Bloechl1994,Kresse1999} adapted pseudopotentials were used.
For Pb and Sn atoms, the $d$ semi-core states were treated explicitly as valence states with an extended version of PAW potentials. 
The generalized gradient approximation (GGA) parametrized by Perdew-Burke-Ernzerhof (PBE)\cite{Perdew1996} was employed for the evaluation of the exchange-correlation functional.
The Brillouin zone was sampled with $\Gamma$-centered \textit{k}-point grids of 5$\times$5$\times$3, 5$\times$5$\times$3, and 5$\times$5$\times$5 for the orthorhombic, tetragonal, and cubic unit cells, respectively.
The minimal cutoff energy was set to 500 eV for both the geometry optimization calculations and the self-consistent energy calculations.
The convergence threshold for self-consistent-field iteration was set to $10^{-4}$ eV.
As a starting point for the geometry optimization, we used experimentally obtained data from powder X-ray diffraction (PXRD) measurements of ${\MA}PbI_3$ carried out by Stoumpos \textit{et al.}\cite{Stoumpos2013b} for the tetragonal phase, and Baikie \textit{et al.}\cite{Baikie2013} for the orthorhombic and cubic phases (with additional theoretical results by Giorgi \textit{et al.}).\cite{Giorgi2013}

\subsection{Thermodynamic model}\label{sec:rsm}

The thermodynamic stability of alloys can often be understood in terms of nearest-neighbor two-particle interactions only, especially if the substitution compounds are chemically similar.
To gain some insight into the thermodynamic stability of mixed \textit{Pb:Sn} perovskites, we describe every $BX_6$ octahedron as one unit placed on a three-dimensional lattice, which interacts only with its $Z=6$ nearest-neighbor octahedra.
Small distance variations in bond length, which occur in structures for orthorhombic and tetragonal phase due to unequal primitive vector lengths can be neglected.
To be specific, we start with the cohesive energies $E$ of various mixed structures obtained from the self-consistent DFT calculations and describe them as a linear combination of the binding energies $\mathcal{E}_{PbPb}$, $\mathcal{E}_{PbSn}$ and $\mathcal{E}_{SnSn}$ between $PbX_6$ and $SnX_6$ octahedra. Thus, we write
\begin{equation}\label{eq:Ebind}
E = \frac{1}{2} \left( N_{PbPb} \, \mathcal{E}_{PbPb} + N_{PbSn} \, \mathcal{E}_{PbSn} + N_{SnSn} \, \mathcal{E}_{SnSn} \right) \; ,
\end{equation}
where $N_{PbPb}$, $N_{PbSn}$, $N_{SnSn}$ denote the number of bonds between the corresponding components and the factor $1/2$ has been added to avoid double-counting of the bonds.
A simple bond-counting argument shows that the number of $PbX_6$ octahedra is $N_{Pb} =  N_{PbPb}/6 + N_{PbSn}/12$ and we define the fractions $n_{Pb (Sn)} = N_{Pb (Sn)} / N$ with respect to the total number $N$ of octahedra in the system. Note that $n_{Pb} + n_{Sn} = 1$.

Equation (\ref{eq:Ebind}) is the starting point for the Regular Solution Model (RSM).\cite{Haasen1996}
Its key quantity is the free energy of mixing $F^M = U^M - T S^M$, which is defined as the difference in the Helmholtz free energy between a non-mixed system ($N_{PbSn} = 0$), and a stochastically intermixed ensemble (each octahedron is filled with $Pb$ with probability $n_{Pb}$).
The internal energy of mixing $U^M$ is the stochastic average of the difference in cohesive energy over $N$ octahedra and takes the form
\begin{equation*}\label{eq:umix}
U^M(n_{Pb}) = N \, n_{Pb} (1 - n_{Pb}) \, 6 \, \varepsilon   \; .
\end{equation*}
The parameter
\begin{equation}\label{eq:epsilon}
\varepsilon = \mathcal{E}_{PbSn} - \frac{\mathcal{E}_{PbPb} + \mathcal{E}_{SnSn}}{2}  \;
\end{equation}
determines whether energy is gained or released during the formation of two heterogeneous $PbX_6 - SnX_6$ bonds by means of breaking homogeneous $PbX_6 - PbX_6$ and $SnX_6 - SnX_6$ bonds.
Within the RSM, the entropy of mixing $S^M$ depends solely on the number of possible configurations and thus on $n_{Pb}$ via the following expression:
\begin{equation}\label{eq:smix}
S^M(n_{Pb}) = N \, k_B \left[ n_{Pb} \ln(n_{Pb}) + (1 - n_{Pb}) \ln (1 - n_{Pb}) \right]  \; ,
\end{equation}
where $k_B$ is the Boltzmann constant.

A mixed state is called \textit{ideal} if $\varepsilon=0$ and \textit{regular} otherwise.
In the case of a regular solution, there are two possible scenarios.
If the mixing parameter $\varepsilon$ is negative, the heterogeneous bond is always energetically favorable and thus mixing of the components is preferred at any temperature.
The more interesting case where the mixing parameter $\varepsilon$ is positive, requires a closer examination.
In this case, the miscibility of the components depends on the interplay between the internal energy of mixing $U^M$, which favors the formation of clusters of only one kind of octahedra and thus phase separation, and the temperature-dependent entropic term, $- T S^M$, which favors mixing.

At low temperatures, the $U^M$ term prevails and the system separates into two phases at all ratios $n_{Pb}$.
With increasing temperature, the entropic term causes a change in the shape of the $F^M(n_{Pb})$ curve, which leads to the reduction of the so-called mixing gap.
At a the \textit{upper critical solution temperature} $T_{UCS}$, this gap closes and the system experiences no phase separation anymore.
Within the RSM, the value of $T_{UCS}$ can be calculated from the condition $0 = \left. \frac{\partial F^M}{\partial n_{Pb}}\right|_{T=T_{UCS}}$ as:
\begin{equation}\label{eq:t_c}
T_{UCS} = \frac{6 \, \varepsilon \, (2 n_{Pb} - 1)}{k_B \left[\ln(n_{Pb}) - \ln(1 - n_{Pb})\right]}   \; ,
\end{equation}
with a maximum $T_{UCS} = 3 \varepsilon / k_B$ at $n_{Pb}=0.5$.
Experimental deviations from this universal result are observed, e.g., for polymer mixtures, where the building blocks are of different volume and geometry.
Such systems are better described by the Flory-Huggins solution model,\cite{Flory1951,Huggins1943} which is an extension of the RSM.

Note that we do not include the entropic contribution stemming from the different orientations of the organics.
This contribution is key for understanding the phase transition between, e.g., the tetragonal and the cubic phase.
However, it can be neglected for the question of mixing within one phase, because such entropic terms cancel out in equation (\ref{eq:smix}).

\section{Geometry of optimized structures}

In this section, we analyze the optimized structures of hybrid metal-organic halide perovskites ${\MA}(Pb:Sn)X_3$.
First, we discuss the geometry of the \textit{non-mixed} structures containing solely either $Pb$ or $Sn$.
In particular, we compare the octahedral distortions in the three phases and among different compositions and provide for a possible explanation within the framework of the orbital energy matching model.\cite{Tricker1978}
We then discuss the \textit{mixed} systems, i.e., those with both $Pb$ and $Sn$ as metal cations.
Understanding the geometry of the mixed systems will be the basis for the thermodynamic considerations in section \ref{sec:energy_thermo}.

\subsection{Comparison of non-mixed hybrid perovskites}\label{sec:geom_nonmix}

\begin{figure}
	\includegraphics[width=0.3\linewidth]{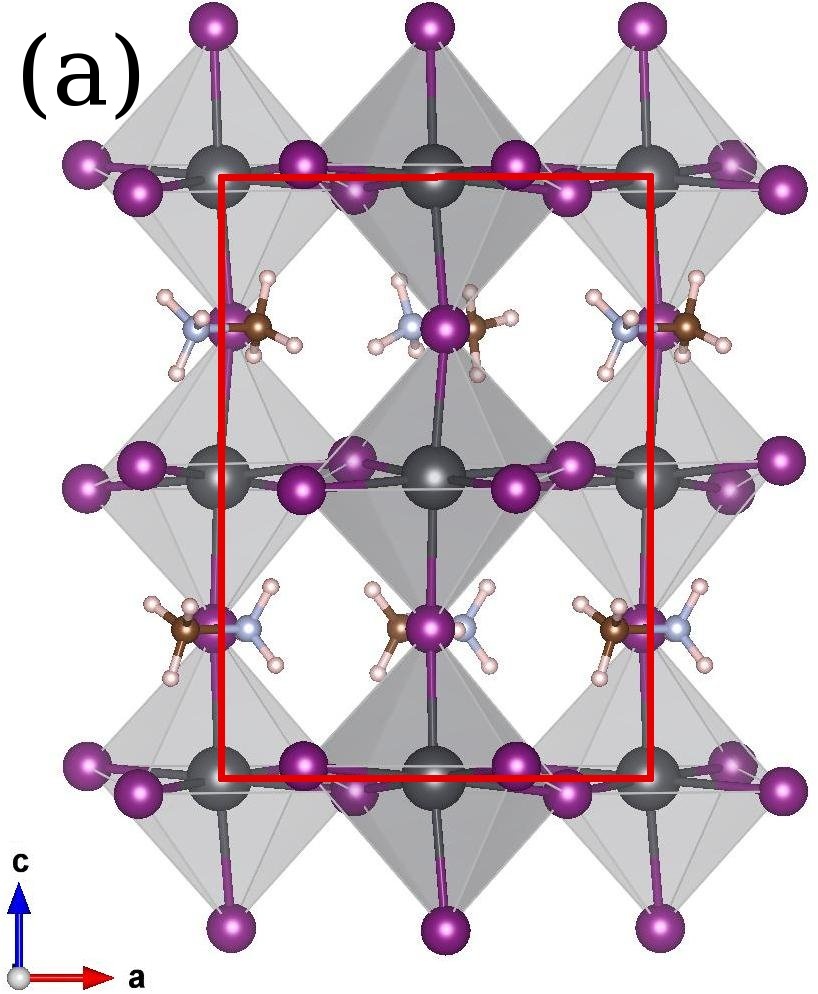}
	\includegraphics[width=0.3\linewidth]{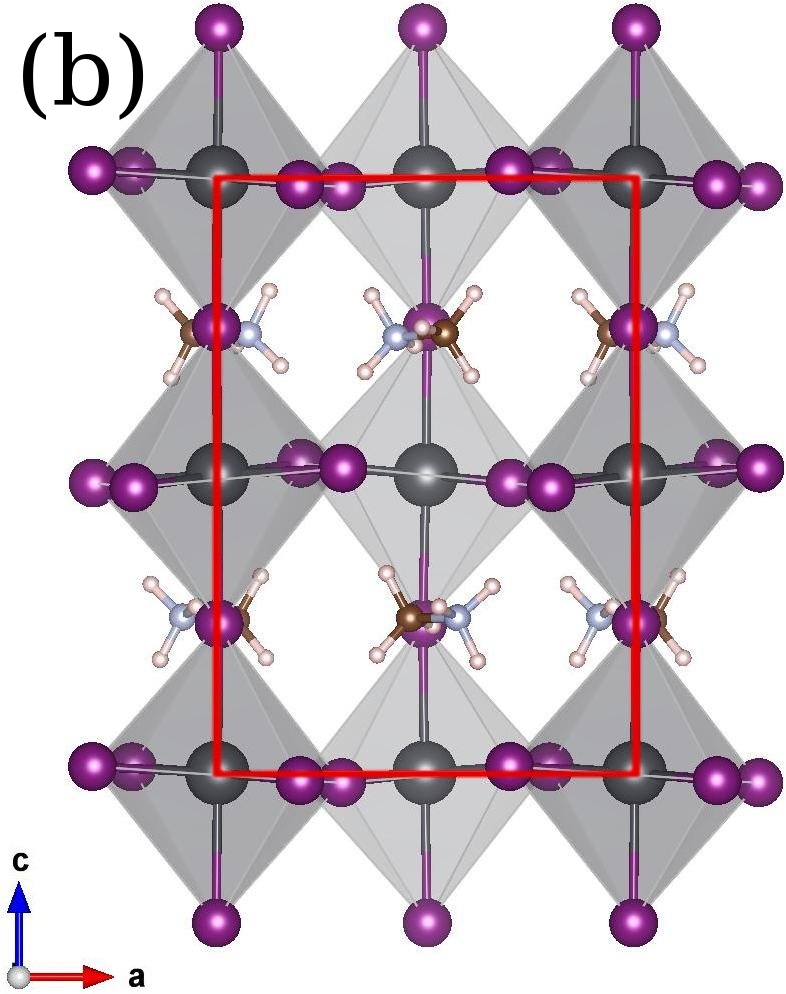}
	\includegraphics[width=0.3\linewidth]{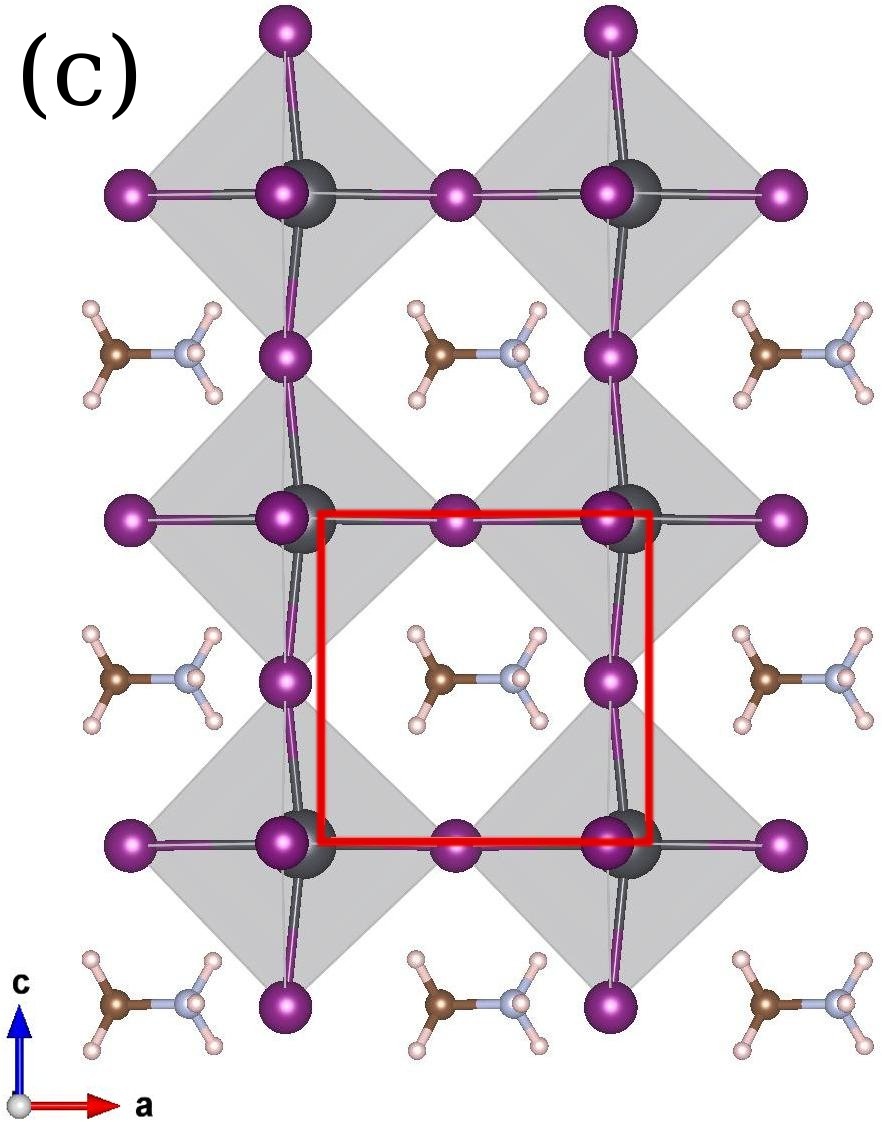}\\
	\includegraphics[width=0.3\linewidth]{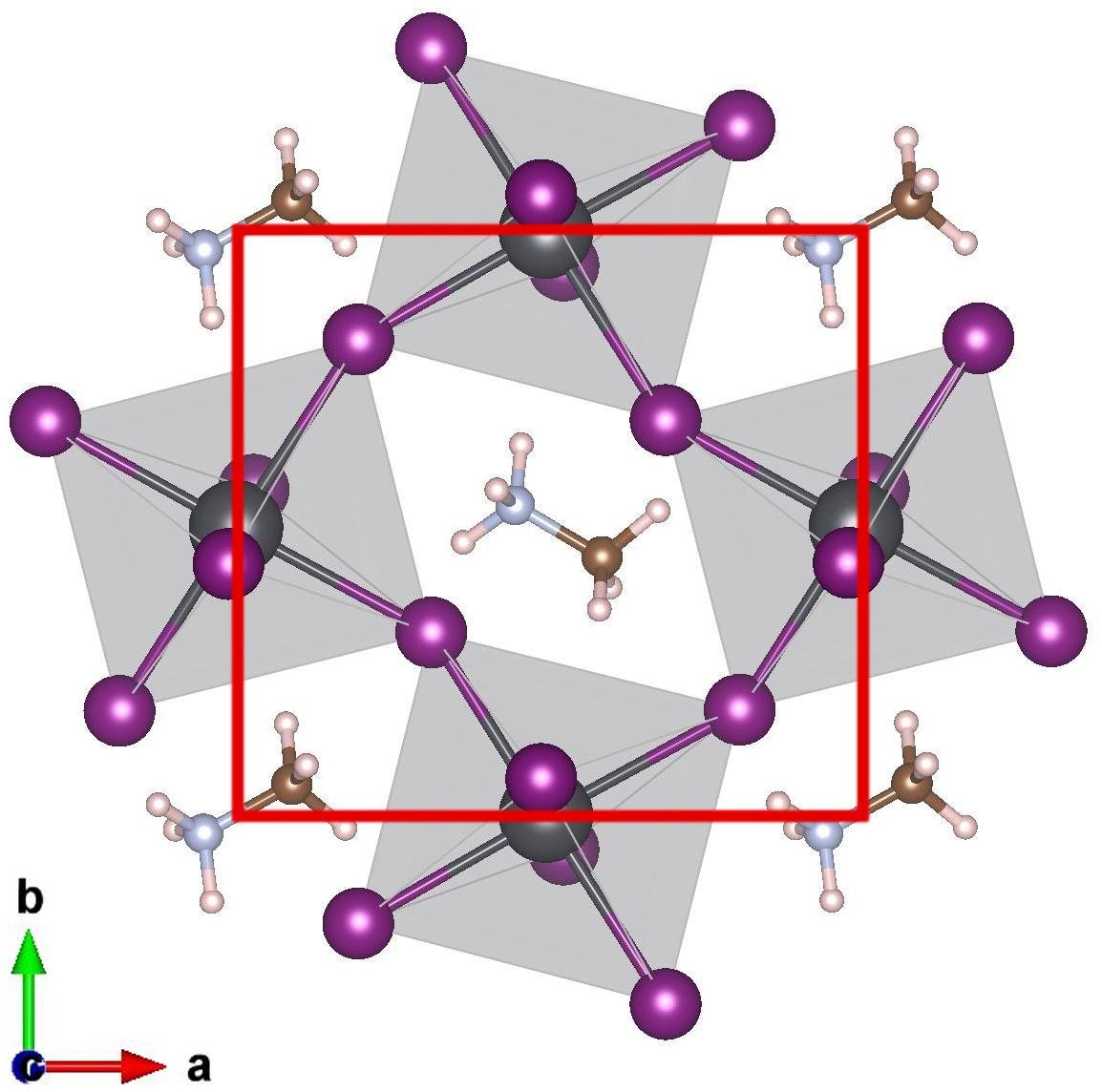}
	\includegraphics[width=0.3\linewidth]{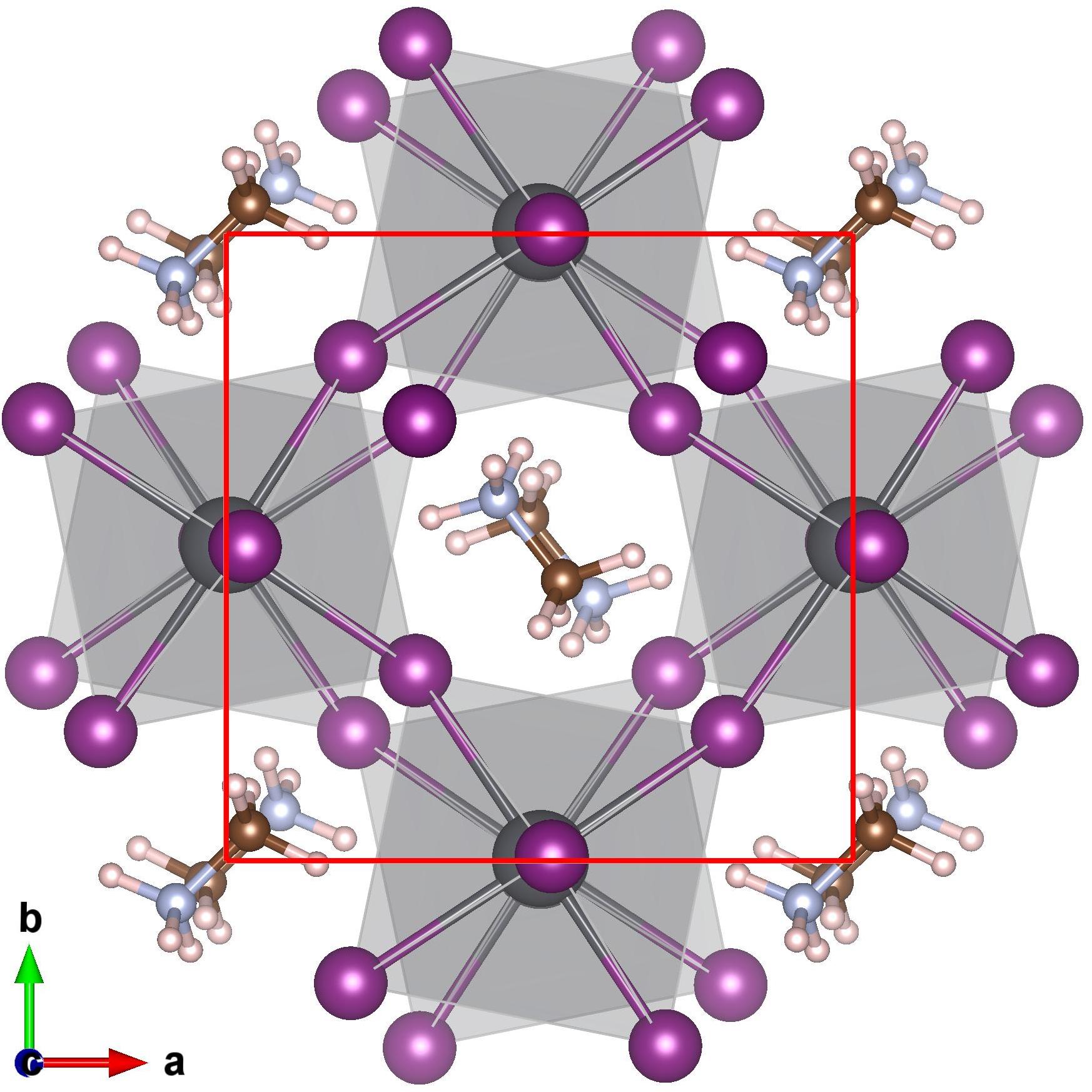}
	\includegraphics[width=0.3\linewidth]{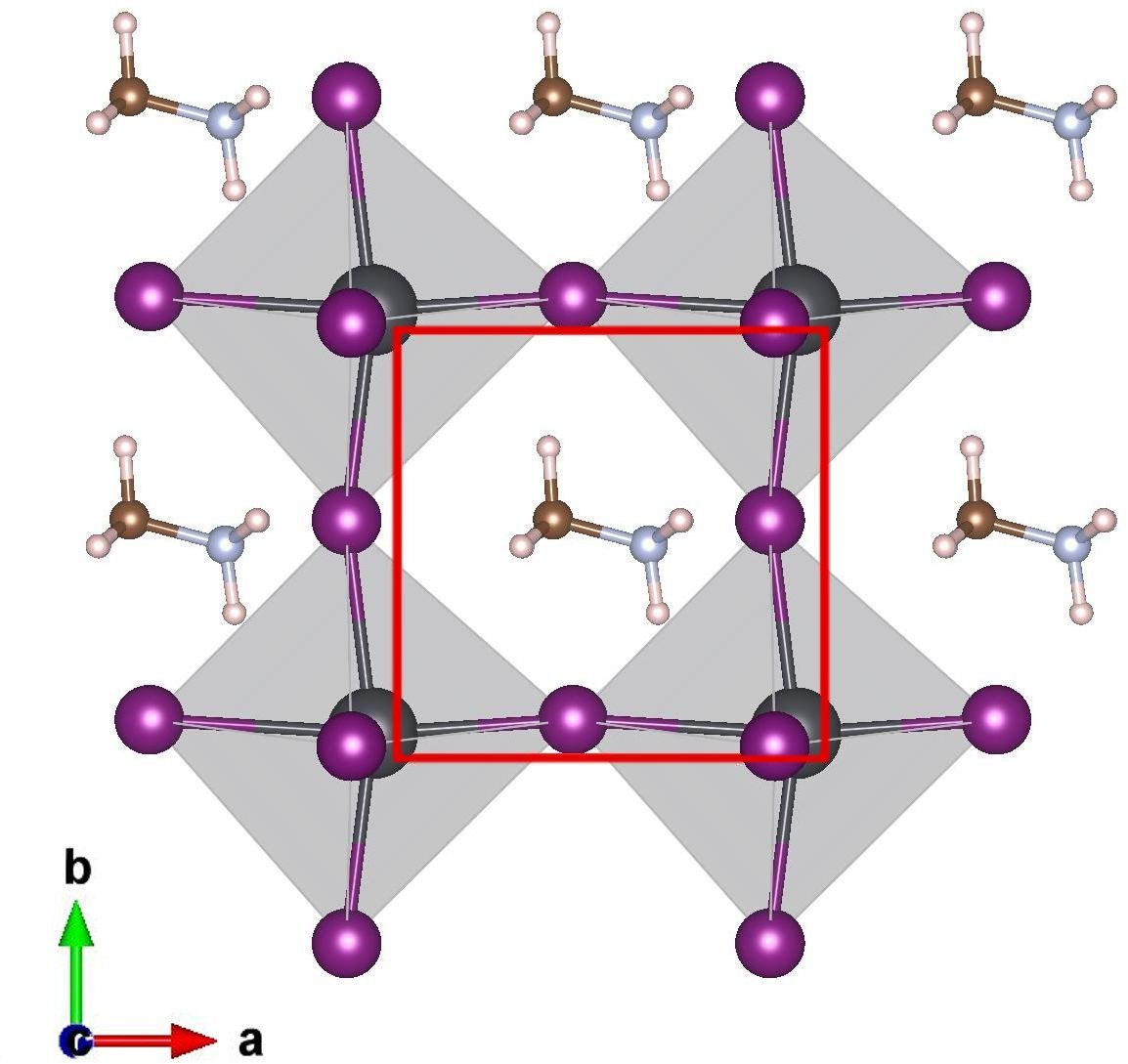}
	\caption{Comparison of (a) orthorhombic, (b) tetragonal and (c) cubic perovskite phases obtained from structural optimization for the case of ${\MA}PbI_3$. Top row: a-c-plane, bottom row: a-b-plane.}
	\label{fig:geom_otc}
\end{figure}

For all investigated hybrid metal-organic perovskites, the lowest energy configuration forms an orthorhombic Bravais lattice, i.e., the lattice vectors differ in length ($a \neq b \neq c$), but are orthogonal.
Concordantly, it is the preferred structure at low temperatures.
However, we also investigated the tetragonal ($a = b \neq c$) and cubic ($a = b = c$) phases, which have been observed in experiments at higher temperatures.\cite{Poglitsch1987}
For the prototypical ${\MA}PbI_3$, the transition temperatures are $\unit[162]{K}$ ($\unit[-111]{^\circ C}$) and $\unit[327]{K}$ ($\unit[54]{^\circ C}$).
The geometry of all three phases is illustrated for a particular example in Figure \ref{fig:geom_otc}.
Note that in all cases the $PbI_3$ octahedra are corner-connected, i.e., each $Pb$ cation has exactly six $I$ anions as nearest neighbors and each $I$ anion is connected to two $Pb$ cations.
In the tetragonal phase, alternating octahedra along the $c$ axis are rotated within the a-b-plane by an angle of approximately $22.5^\circ$.
In the orthorhombic phase, the optimal (lowest energy) orientation of the organics is predetermined by the nearly cuboctahedral cavities between the $PbI_6$ octahedra and any rotation of the ${\MA}$ cations is geometrically hindered.
As a consequence, the equilibrium volume of the cubic unit cell of ${\MA}PbI_3$ is 5\% larger than the orthorhombic cell.
The cubic symmetry allows at least three different equivalent orientations (due to the three dimensions).
Consequently, rotation of MA is more likely in the cubic phase. We found an upper limit for the rotational barrier for MA of approximately $\unit[60]{meV}$ using minimal-energy-chain-of-states calculations.
Even though the finite temperature dynamics of the organics in cubic ${\MA}PbI_3$ is an interesting research topic of its own,\cite{Frost2014} in this paper we will only consider the lowest energy configuration in the cubic phase, where all organic groups are aligned parallelly (shown in Figure \ref{fig:geom_otc}c).
Some details of the calculations are worthwhile to be noted explicitly:
The unit cells of the orthorhombic and tetragonal phases contain 4 formula units of ${\MA}PbI_3$ each.
While the rotation of the octahedra is fixed in the orthorhombic and tetragonal phases, it is necessary to use an even number of primitive cells in the cubic supercell in each space direction to allow for possible alternating rotations of the octahedra.
Thus, for the cubic phase, we used a larger supercell of 2{\texttimes}2{\texttimes}2 primitive cells, thus containing 8 formula units.
%However, we found that the octahedra in the cubic phase do not rotate during relaxation.
While the octahedra deform during relaxation (as discussed below), we do not find any alternating rotations of the octahedra in the cubic phase.
For the orthorhombic phase, we allowed a full relaxation of both ion positions and lattice vectors (including volume).
In the tetragonal and cubic phases, we found the optimal volume by a fit to the Murnaghan equation of state\cite{Murnaghan1944}, where we preserved the tetragonal symmetry of the unit cell by keeping the $c/a$-ratio at the experimentally obtained value of 1.429 \cite{Stoumpos2013b}.

\begin{table*}[ht!]
	\begin{tabular}{l|ccc|cc|c}
		\hline
		& \multicolumn{3}{c|}{\textbf{orthorhombic}} & \multicolumn{2}{c|}{\textbf{tetragonal}} & \textbf{cubic} \\
		\textbf{compound}  & \textbf{a[\AA]} & \textbf{b[\AA]} & \textbf{c[\AA]} & \textbf{a[\AA]} & \textbf{c[\AA]} & \textbf{a[\AA]} \\
		\hline
		${\MA}PbI_3$                    &       9.18    &   8.65  &  12.87  &  9.06  &  12.94  &  6.45  \\  
		${\MA}Sn_{0.25}Pb_{0.75}I_3$    &       9.08    &   8.66  &  12.85  &  9.03  &  12.90  &  6.43  \\  
		${\MA}Sn_{0.5}Pb_{0.5}I_3$[c]   &       9.05    &   8.64  &  12.79  &  9.03  &  12.90  &  6.44  \\  
		${\MA}Sn_{0.5}Pb_{0.5}I_3$[d]   &       9.18    &   8.58  &  12.78  &  9.01  &  12.87  &  6.43  \\  
		${\MA}Sn_{0.5}Pb_{0.5}I_3$[l]   &       9.16    &   8.63  &  12.79  &  9.04  &  12.91  &  6.42  \\  
		${\MA}Sn_{0.75}Pb_{0.25}I_3$    &       9.04    &   8.61  &  12.73  &  9.00  &  12.85  &  6.41  \\  
		${\MA}SnI_3$                    &       9.02    &   8.64  &  12.66  &  8.84  &  12.54  &  6.40  \\  
		\hline
		${\MA}PbBr_3$                   &       8.74    &   8.11  &  12.10  &  8.54  &  12.21  &  6.04  \\  
		${\MA}Sn_{0.25}Pb_{0.75}Br_3$   &       8.64    &   8.12  &  12.05  &  8.52  &  12.17  &  6.04  \\  
		${\MA}Sn_{0.5}Pb_{0.5}Br_3$[c]  &       8.62    &   8.09  &  12.06  &  8.52  &  12.17  &  6.04  \\  
		${\MA}Sn_{0.5}Pb_{0.5}Br_3$[d]  &       8.72    &   8.09  &  12.01  &  8.50  &  12.14  &  6.04  \\  
		${\MA}Sn_{0.5}Pb_{0.5}Br_3$[l]  &       8.65    &   8.08  &  12.09  &  8.52  &  12.18  &  6.04  \\  
		${\MA}Sn_{0.75}Pb_{0.25}Br_3$   &       8.67    &   8.12  &  12.00  &  8.49  &  12.12  &  6.04  \\  
		${\MA}SnBr_3$                   &       8.91    &   8.11  &  12.04  &  8.34  &  11.83  &  6.04  \\  
		\hline
		${\MA}PbCl_3$                   &       8.47    &   7.62  &  11.58  &  8.23  &  11.76  &  5.76  \\  
		${\MA}Sn_{0.25}Pb_{0.75}Cl_3$   &       8.35    &   7.76  &  11.53  &  8.21  &  11.73  &  5.76  \\  
		${\MA}Sn_{0.5}Pb_{0.5}Cl_3$[c]  &       8.30    &   7.69  &  11.64  &  8.21  &  11.73  &  5.76  \\  
		${\MA}Sn_{0.5}Pb_{0.5}Cl_3$[d]  &       8.37    &   7.74  &  11.50  &  8.19  &  11.70  &  5.76  \\  
		${\MA}Sn_{0.5}Pb_{0.5}Cl_3$[l]  &       8.43    &   7.72  &  11.47  &  8.21  &  11.74  &  5.76  \\  
		${\MA}Sn_{0.75}Pb_{0.25}Cl_3$   &       8.45    &   7.79  &  11.58  &  8.18  &  11.68  &  5.76  \\  
		${\MA}SnCl_3$                   &       8.74    &   7.93  &  11.77  &  8.04  &  11.40  &  5.76  \\ 
		\hline
	\end{tabular}
	\caption{Calculated crystallographic data of mixed \textit{Pb:Sn} hybrid perovskites in the orthorhombic ($a \neq b \neq c$), tetragonal ($a = b \neq c$) and cubic ($a = b = c$) phases.
		The suffixes [c], [d] and [l] refer to the chain, diagonal and layered structures, respectively (see Figure \ref{fig:mixed_cdl} below).
		Note that the volume per formula is $V_0 = a^3$ for the cubic cells, but $V_0 = a \, b \, c / 4$ and $V_0 = a^2 \, c / 4$ for the orthorhombic and tetragonal cells, respectively.
	}
	\label{tab:dat_abc}
\end{table*}

The structural parameters calculated  for both mixed and non-mixed structures are listed in Table \ref{tab:dat_abc}.
Already for the non-mixed structures, we see that the substitution of the halogen has a much stronger influence on the crystallographic data than the substitution of the metal cation.
To give an example, the lattice parameter $c$ in the orthorhombic phase changes only from $\unit[12.87]{\AA}$ for ${\MA}PbI_3$ to $\unit[12.66]{\AA}$ for ${\MA}SnI_3$, while already the bromide-containing structure ${\MA}PbBr_3$ has a value for $c = \unit[12.10]{\AA}$ well below both iodide compounds.
The value for $c$ in the chloride structure is even lower.
Table \ref{tab:dat_abc} shows that this holds for each of the investigated ${\MA}BX_3$ structures.

In summary, the volume of the unit cell is the largest in the (high-temperature) cubic phase and the smallest in the (low-temperature) orthorhombic phase and decreases upon increasing electronegativity of the substituted halogen within each phase.

\subsection{Octahedral distortions and tilt angles}\label{sec:octadeform}

\begin{figure}
	\includegraphics[width=0.48\linewidth]{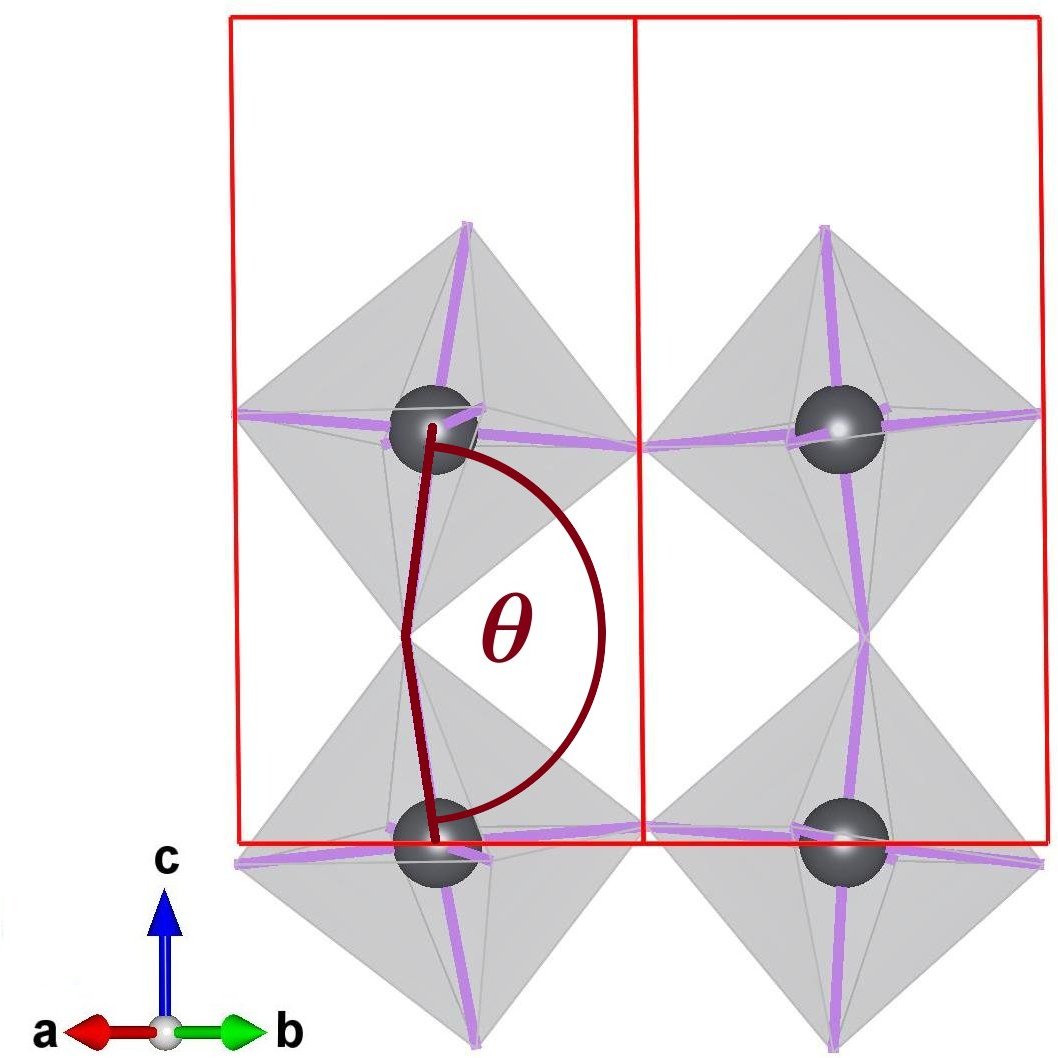}
	\includegraphics[width=0.48\linewidth]{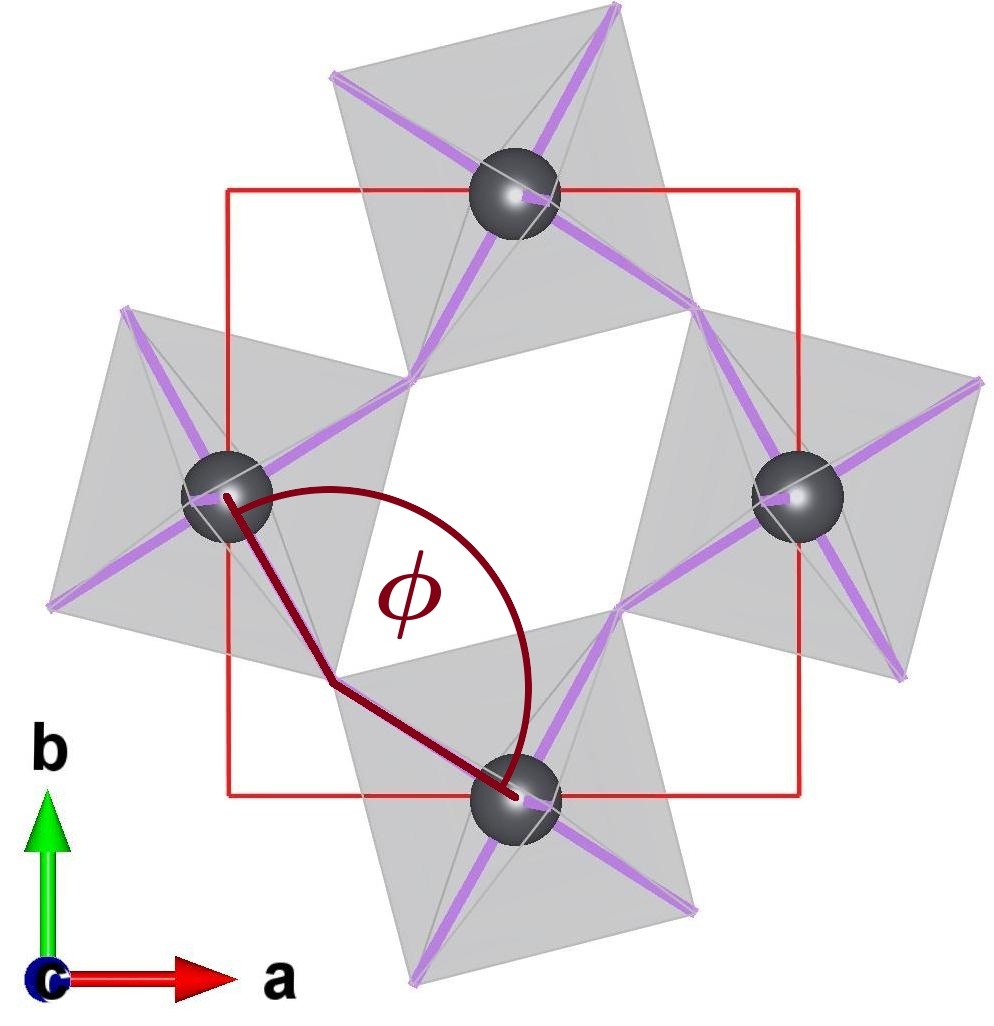}\\
	\includegraphics[width=0.48\linewidth]{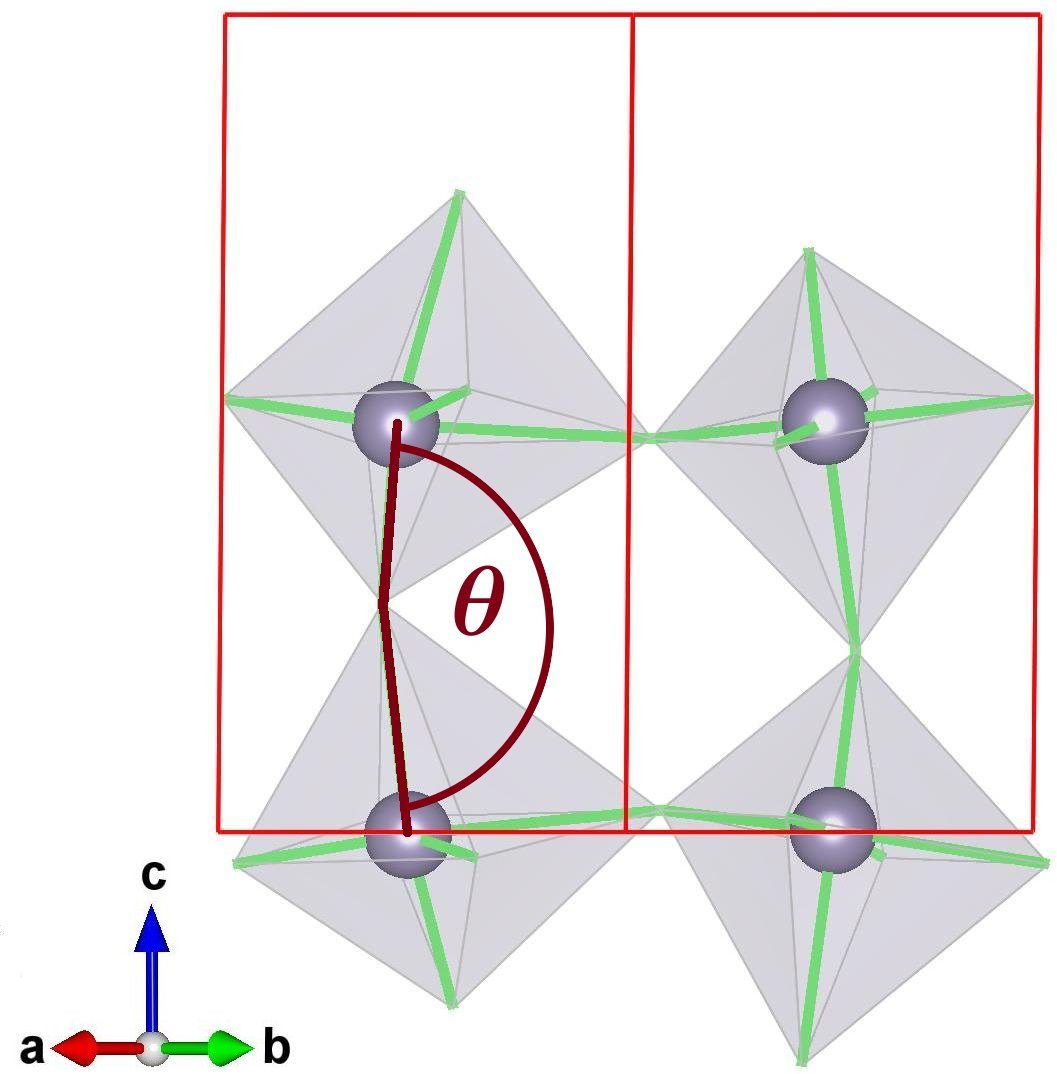}
	\includegraphics[width=0.48\linewidth]{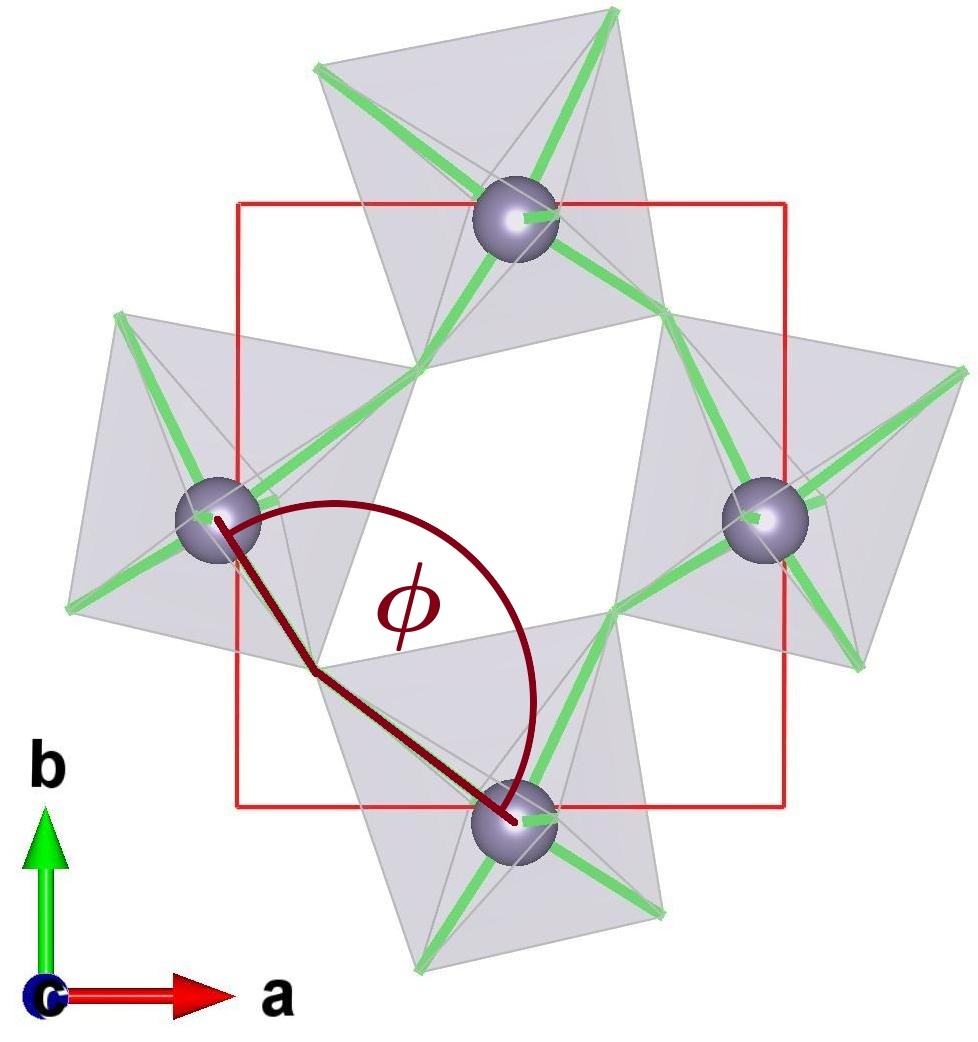}
	\caption{Tilt angles for octahedral distortions. Top row: $MAPbI_3$. Bottom row: $MASnCl_3$. Note that the $Pb$ ion is positioned approximately at the center of the $PbI_3$ octahedra, whereas the $SnCl_3$ octahedra are strongly deformed and $Sn$ is placed eccentrically.}
	\label{fig:angles}
\end{figure}

While $PbI_6$ octahedra keep their regular shape after geometry relaxation, we noticed that they become more deformed in other $BX_6$ compositions.
In order to quantify this observation and investigate the influence of the halogen, we calculate the mean volume of the irregular octahedra, $V_\boxtimes$, and a ratio $\xi$ describing the deformation of the $BX_6$ octahedra.
The latter is defined in terms of the shortest ($d^{B-X}_{min}$) and the longest ($d^{B-X}_{max}$) \textit{B-X} bond lengths as
\begin{equation}\label{eq:rho}
\xi := \frac{d^{B-X}_{max}}{d^{B-X}_{min}} - 1 \;.
\end{equation}
If the octahedron is symmetric and thus all bond length are equal, then $\xi = 0$.
If the octahedron is deformed or if the central metal atom is displaced off the center, then $\xi > 0$.
Furthermore, we calculate two tilt angles, $\phi$ and $\theta$, corresponding to the in-plane and the axial angles of the \textit{X-B-X} bonds, respectively.
A graphical representation of the tilt angles can be found in Figure \ref{fig:angles}, which also shows the extreme cases of octahedral distortion found in the orthorhombic phase.
While the $PbI_3$ octahedra (top row) show only minimal distortions and maintain a regular shape, the $SnCl_3$ octahedra (bottom row) are highly distorted.

In Table \ref{tab:dat_octahedra}, we summarize $\theta$, $\phi$, $\xi$ and $V_\boxtimes$ for all investigated ${\MA}BX_3$ structures.
The tilt angles $\theta$ and $\phi$ strongly depend on whether the system is orthorhombic, tetragonal or cubic, but show smaller variance for different compositions in the same phase.
Note that even in the cubic phase, $\theta$ and $\phi$ deviate from $180^\circ$, which is already visible in Figure \ref{fig:geom_otc}(c) and can be attributed to effects like the interaction of the octahedra with the dipole moment of the ${\MA}$ cation and the sterically hindered rotation of the octahedra\cite{Woodward1997,Garcia-Fernandez2010,Frost2014}.

The mean volume of the octahedra $V_\boxtimes$, on the other hand, shows a strong dependence on the halogen, but depends only weakly on the type of the Bravais lattice.
Note that the octahedra always fill between 16.5\% (cubic ${\MA}PbI3_3$) and 18.2\% (orthorhombic ${\MA}SnCl_3$) of the unit cell.
In the iodide structures, the $PbI_6$ octahedra have larger $V_\boxtimes$ than the $SnI_6$ octahedra in all phases.
Such a general statement can not be made in the case of bromide and chloride containing structures:
In the orthorhombic phase, the tin bromide and tin chloride octahedra are larger than their lead counterparts; while in the tetragonal phase, the lead bromide and lead chloride octahedra are larger.
Comparing $V_\boxtimes$ for the same composition along different phases, we see that $V_\boxtimes$ is maximal in the tetragonal phase for $Pb$-centered octahedra, but minimal for $Sn$-centered ones.

In all phases, the deformation parameter $\xi$ is much larger for $Sn$-derived structures than for their $Pb$ counterparts.
Among them, $SnCl_3$ (in the low-temperature phases) exhibits the strongest deformations.
This can be attributed to the high electronegativity of the ligand and can be explained within the framework of the ``orbital energy matching'' model.\cite{Tricker1978}
It is based on the Linear Combination of Atomic Orbitals approximation (LCAO).
According to this model, the formation of a strong bond between the orbital of the metal cation (tin) and the halogen $np$ orbital occurs when the orbitals are of compatible symmetry and similar energy.
To match the energy of a $np$ halogen $p$-valence orbital, the percentage of $5s$ and $5p$ atomic tin orbitals is changed during the hybridization.
For tin iodide structures, the percentage of the $p$ orbital in the bonding hybrid tin orbital is maximal, since the $5p$ orbital of iodine has the highest cohesive energy among the discussed halogens.
Because of the orthogonality requirement, the non-bonding hybrid orbital containing the lone pair has mainly $s$-character and low spatial directionality.
With the decreasing cohesive energy of the $np$ orbital of the halogen, the non-bonding hybrid orbital of tin obtains stronger $p$ character, which makes it spatially directed and more stereochemically active.
Thus, the octahedral distortions caused by the stereochemically active lone pair on tin increases upon the increase of the electro-negativity of the involved halogen.
Apparently, our plane-wave DFT results confirm this simple LCAO-based argument.

\begin{table*}
	\begin{tabular}{l|cccc|cccc|cccc}
		\hline
		& \multicolumn{4}{c|}{\textbf{orthorhombic}} & \multicolumn{4}{c|}{\textbf{tetragonal}} & \multicolumn{4}{c}{\textbf{cubic}} \\
		& $\bm{\theta}[^\circ]$ & $\bm{\phi}[^\circ]$ & \textbf{$\bm{V}_\boxtimes$[\AA$^3$]} & $\bm{\xi}$
		& $\bm{\theta}[^\circ]$ & $\bm{\phi}[^\circ]$ & \textbf{$\bm{V}_\boxtimes$[\AA$^3$]} & $\bm{\xi}$
		& $\bm{\theta}[^\circ]$ & $\bm{\phi}[^\circ]$ & \textbf{$\bm{V}_\boxtimes$[\AA$^3$]} & $\bm{\xi}$ \\
		\hline
		${\MA}PbI_3$  & 161 & 148 & 42.22 &  3\%   & 174 & 157 & 46.06 &  2\%    & 172 & 168 & 44.17 & 8\% \\
		${\MA}SnI_3$  & 164 & 150 & 43.93 &  9\%   & 176 & 162 & 41.87 &  2\%    & 173 & 167 & 43.71 & 16\% \\
		%\hline
		${\MA}PbBr_3$ & 159 & 153 & 38.28 &  2\%   & 173 & 159 & 38.93 &  4\%    & 173 & 168 & 36.73 & 7\% \\
		${\MA}SnBr_3$ & 162 & 146 & 38.86 & 26\%   & 176 & 163 & 34.97 &  9\%    & 172 & 168 & 36.73 & 22\% \\
		%\hline
		${\MA}PbCl_3$ & 159 & 155 & 33.21 &  3\%   & 173 & 165 & 33.47 &  6\%    & 172 & 168 & 31.91 & 4\% \\
		${\MA}SnCl_3$ & 162 & 143 & 37.10 & 42\%   & 173 & 159 & 31.23 & 21\%    & 172 & 168 & 31.91 & 24\% \\
		\hline
	\end{tabular}
	\caption{Calculated data for octahedra in different phases and with different compositions. $V_\boxtimes$ denotes the mean octahedral volume. For the definition of tilt angles $\theta$ and $\phi$, see Figure \ref{fig:angles}. The deformation ratio $\xi$ is defined in equation (\ref{eq:rho}).}
	\label{tab:dat_octahedra}
\end{table*}

\subsection{Properties of mixed \textit{Sn:Pb} hybrid perovskites}

\begin{figure}
	\includegraphics[width=0.48\linewidth]{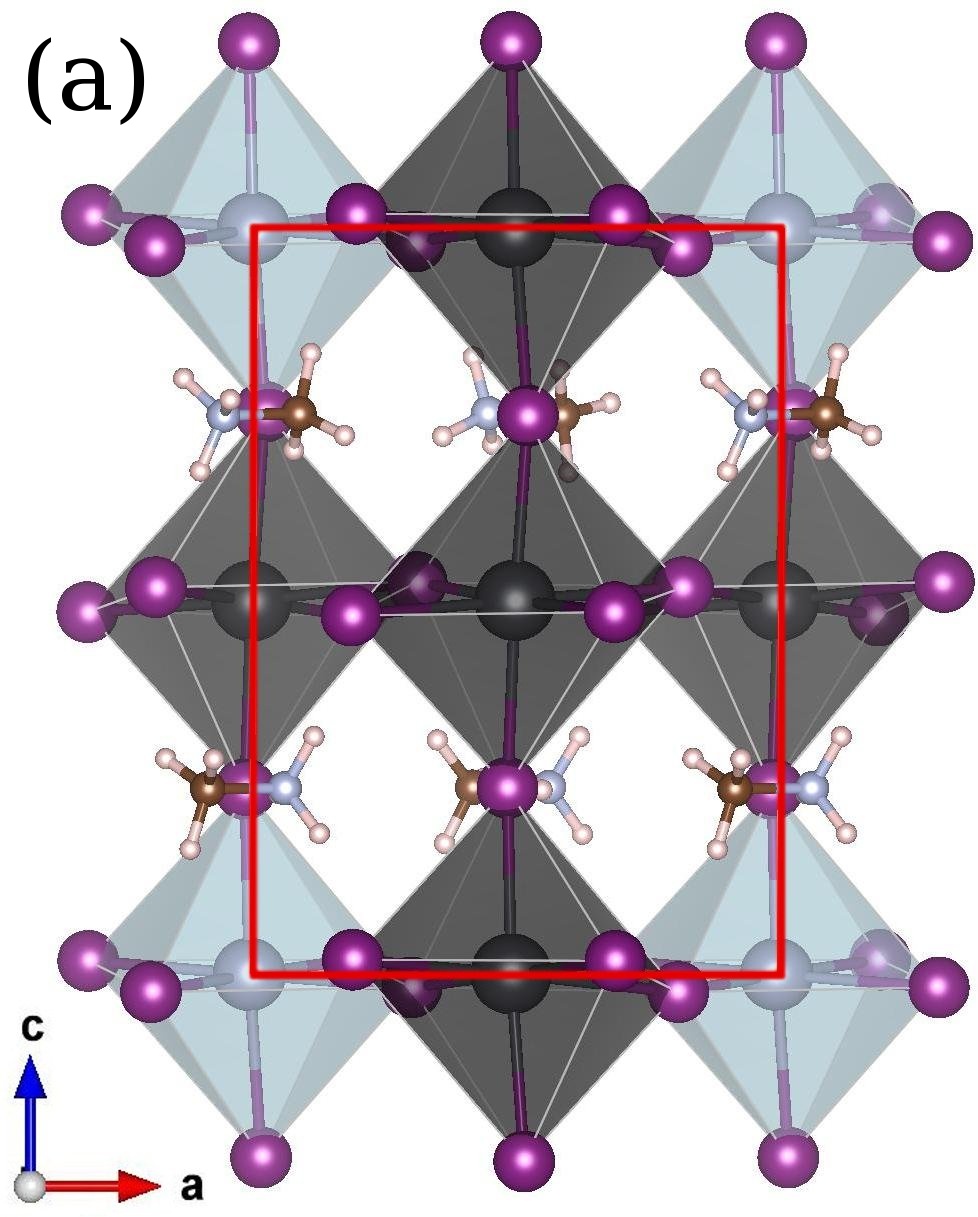}
	\includegraphics[width=0.48\linewidth]{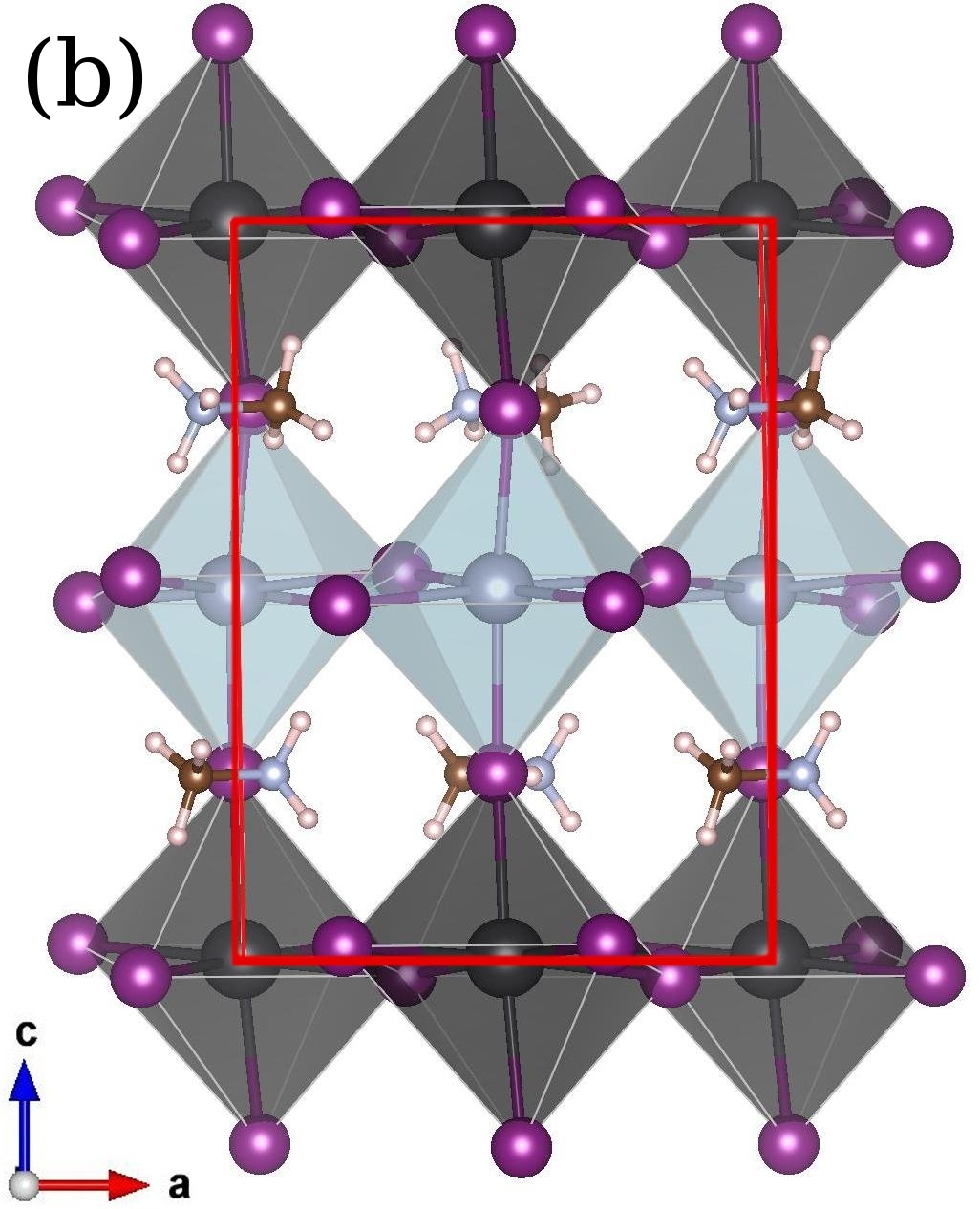}\\
	\includegraphics[width=0.48\linewidth]{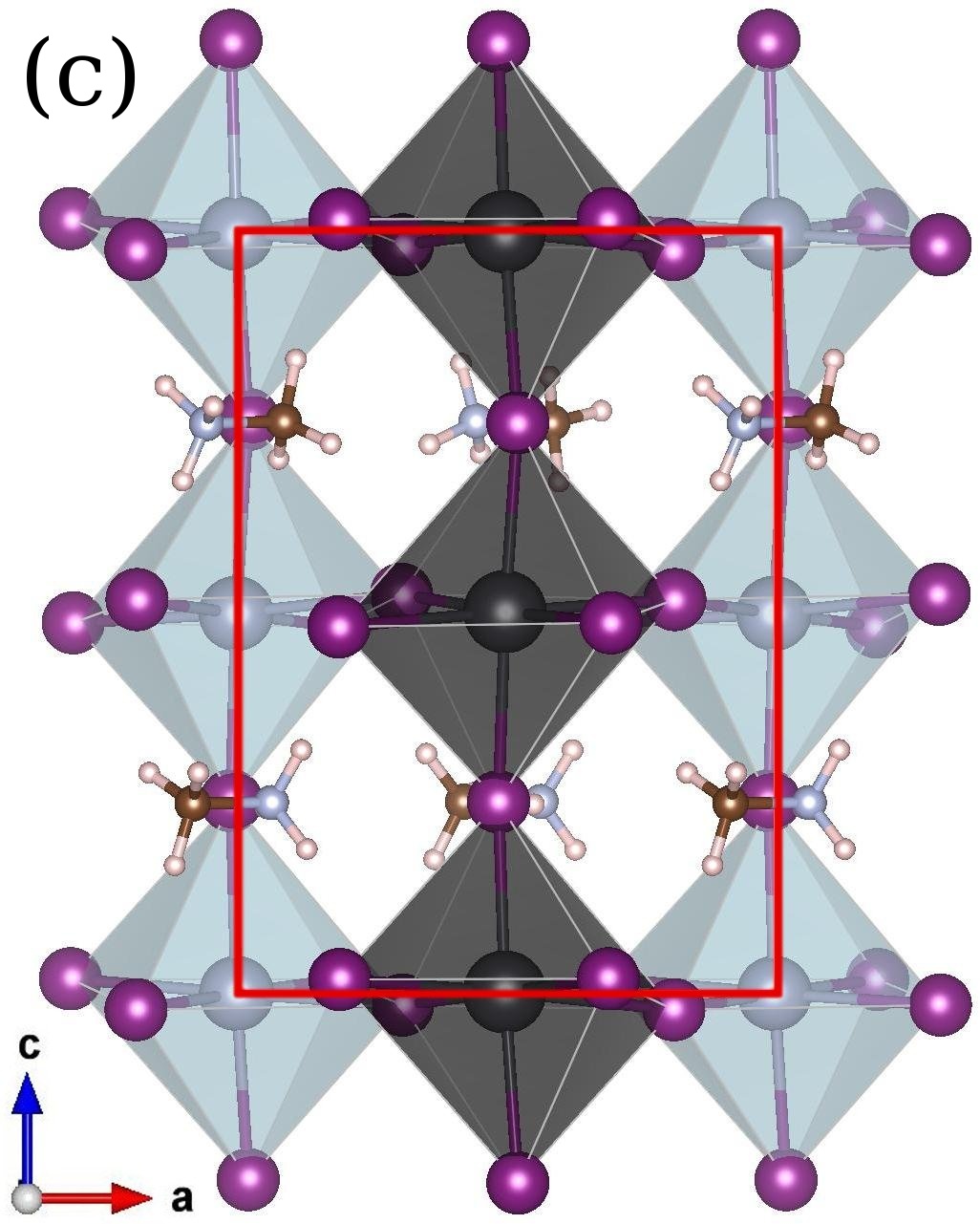}
	\includegraphics[width=0.48\linewidth]{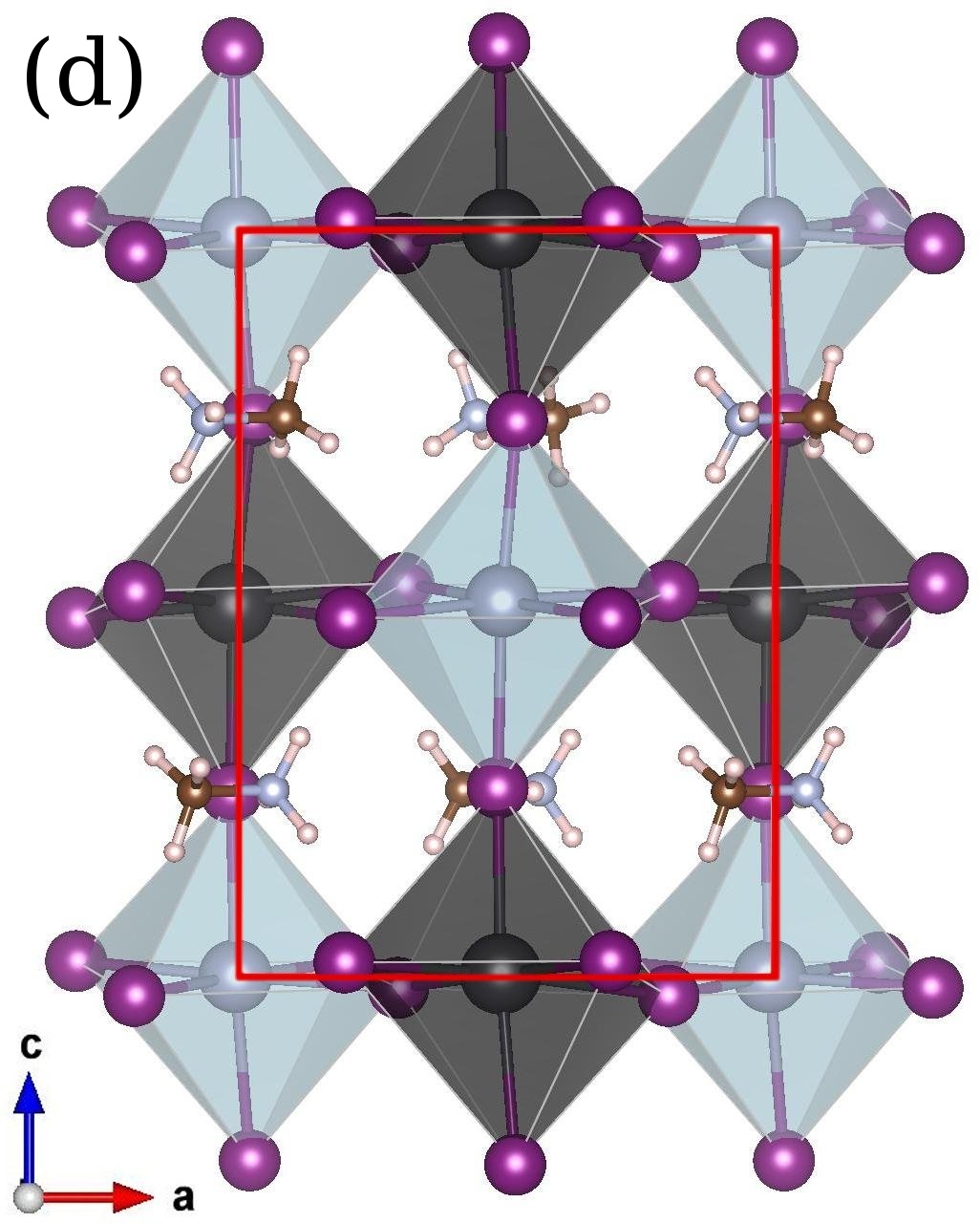}
	\caption{Mixed orthorhombic \textit{Pb:Sn} perovskite structures with the following ratios and composition: (a) 3:1, (b) 1:1 and ``layer'', (c) 1:1 and ``chain'', (d) 1:1 and ``diagonal''.
		Dark octahedra are Pb-centered, light octahedra are Sn-centered.}
	\label{fig:mixed_cdl}
\end{figure}

To investigate mixed \textit{Pb:Sn} hybrid perovskites, we calculated cohesive energies and optimal geometries of five different unit cells as shown in Figure \ref{fig:mixed_cdl}.
While the geometry will be discussed in this section, results based on the energy value are given in the next section.
Figure \ref{fig:mixed_cdl}(a) shows a structure with a \textit{Pb:Sn}-ratio of \textit{3:1}, corresponding to the formula ${\MA}Sn_{0.25}Pb_{0.75}X_3$.
Similarly, one can construct a cell with a ratio of \textit{1:3} (not shown), corresponding to the formula ${\MA}Sn_{0.75}Pb_{0.25}X_3$.
For the \textit{1:1} ratio (${\MA}Sn_{0.5}Pb_{0.5}X_3$), we considered three different structures that we call ``layer'', ``chain'' and ``diagonal'' as depicted in Figure \ref{fig:mixed_cdl} (b), (c) and (d), respectively.
The ``chains'' are formed by corner-connected octahedra of the same kind along the longest axis ($c$), whereas the ``layered'' structure consists of alternating layers of $Pb$ and $Sn$ centered octahedra, perpendicular to the longest axis.
This also coincides with the different orientation of the MA relative to the chains and layers, respectively.
In the ``diagonal'' structure, shown in Figure \ref{fig:mixed_cdl}(d), all bonds between next nearest octahedra have different metal cations, analogous to a three dimensional checkerboard.
% This is possible due to the setup of the unit cell as shown in Figure \ref{fig:geom_otc}.
The octahedra coordination numbers of all considered structures are listed in Table \ref{tab:Ncoord}.

The structures shown in Figure \ref{fig:mixed_cdl} have been completely relaxed.
It is worthwhile to note two details.
First, the organic MA molecules change their orientations only slightly and approximately maintain their center position.
We conclude that the organic is only weakly influenced by the change of the metal cation.
Second, the octahedra of consecutive layers are slightly tilted out of a-b-plane in opposite directions in all 1:1 structures, which can be seen, e.g., from the positions of the left-most iodide ions in Figure \ref{fig:mixed_cdl}.

After relaxing the non-mixed Pb and Sn structures in the cubic phase, we found that the equilibrium lattice constant for chloride structures was $a = \unit[5.763]{\AA}$ for both ${\MA}PbCl_3$ and ${\MA}SnCl_3$.
Likewise for bromide structures, we found $a = \unit[6.040]{\AA}$ for both ${\MA}PbBr_3$ and ${\MA}SnBr_3$ within the accuracy for three decimal digits.
Therefore, we did no further volume optimization for the mixed cubic chloride and bromide structures and only relaxed the atomic positions.
For the iodide structures, however, we found an equilibrium lattice constant of $\unit[6.40]{\AA}$ and $\unit[6.45]{\AA}$ in case of ${\MA}SnI_3$ and ${\MA}PbI_3$, respectively.
Thus, we relaxed all mixed cubic iodide structures with respect to both the volume and the internal degrees of freedom.
In all other cases, we performed a full relaxation as described in section \ref{sec:geom_nonmix}.
The structural results for all phases are summarized in Table \ref{tab:dat_abc}.

Generally, the lattice parameters $a$, $b$ and $c$ of the mixed \textit{Pb:Sn} structures lie between the non-mixed reference structures.
Similar to the non-mixed structures, the halogen have stronger influence on the crystallographic data than the substitution of metal ions:
Within each of the three phases (orthorhombic, tetragonal, cubic), all lattice parameters $a$, $b$ and $c$ of the mixed \textit{Pb:Sn} structures lie in non-overlapping intervals for the iodide, bromide and chloride compounds, respectively.
Overall, the cubic phase shows the smallest relative difference in the crystallographic data.

The tilt angles $\phi$ and $\theta$ of all mixed structures tend to take the average of the non-mixed ones for all octahedra.
This can be easily explained by the fact that their rotation is sterically hindered by neighboring octahedra.
Analyzing the distribution of $V_\boxtimes$ and $\xi$ of the octahedra in mixed systems, we find that the individual octahedra maintain their non-mixed characteristics with only small variance.
Thus, it is a good approximation to think of the $PbX_6$ and $SnX_6$ as rigid building blocks of the mixed \textit{Pb:Sn} structures.

\begin{table}[h]
	\begin{tabular}{lccc}
		\hline
		\textbf{compound} & $\bm{N_{PbPb}}$ & $\bm{N_{PbSn}}$ & $\bm{N_{SnSn}}$ \\
		\hline
		${\MA}PbX_3$                   & 6 & 0 & 0 \\
		${\MA}Sn_{0.25}Pb_{0.75}X_3$   & 3 & 3 & 0 \\
		${\MA}Sn_{0.5}Pb_{0.5}X_3$ [c] & 1 & 4 & 1 \\
		${\MA}Sn_{0.5}Pb_{0.5}X_3$ [d] & 0 & 6 & 0 \\
		${\MA}Sn_{0.5}Pb_{0.5}X_3$ [l] & 2 & 2 & 2 \\
		${\MA}Sn_{0.75}Pb_{0.25}X_3$   & 0 & 3 & 3 \\
		${\MA}SnX_3$                   & 0 & 0 & 6 \\
		\hline
	\end{tabular}
	\caption{Octahedra coordination numbers $N$, common to all three phases.
		The suffixes [c], [d] and [l] correspond to the chain, diagonal and layered structure, respectively (see Figure \ref{fig:mixed_cdl}).
		The mole-fraction $x$ in formula ${\MA}Sn_{1-x}Pb_{x}X_3$ can be calculated from $x = N_{PbPb}/6 + N_{PbSn}/12$.}
	\label{tab:Ncoord}
\end{table}

\section{Energy and thermodynamics} \label{sec:energy_thermo}
\subsection{Cohesive energy results}

\begin{table}
	\begin{tabular}{lccc}
		\hline
		\textbf{compound} & \textbf{X=I} & \textbf{X=Br} & \textbf{X=Cl}\\
		\hline
		${\MA}PbX_3$                   & -50.9020 & -52.7227 & -54.3526\\
		${\MA}Sn_{0.25}Pb_{0.75}X_3$   & -50.8675 & -52.6618 & -54.2971\\
		${\MA}Sn_{0.5}Pb_{0.5}X_3$ [c] & -50.8376 & -52.6278 & -54.2633\\
		${\MA}Sn_{0.5}Pb_{0.5}X_3$ [d] & -50.8512 & -52.6397 & -54.2754\\
		${\MA}Sn_{0.5}Pb_{0.5}X_3$ [l] & -50.8500 & -52.6558 & -54.2806\\
		${\MA}Sn_{0.75}Pb_{0.25}X_3$   & -50.8136 & -52.6035 & -54.2497\\
		${\MA}SnX_3$                   & -50.7635 & -52.6062 & -54.2714 \\
		\hline
	\end{tabular}
	\caption{Cohesive energies $E$ per formula unit in eV for the \textit{orthorhombic} phase calculated using DFT.
		The unit cell used for calculation and shown in Figure \ref{fig:mixed_cdl} contains 4 formula units.}
	\label{tab:Ebind_o}
\end{table}

\begin{table}
	\begin{tabular}{lccc}
		\hline
		\textbf{compound} & \textbf{X=I} & \textbf{X=Br} & \textbf{X=Cl}\\
		\hline
		${\MA}PbX_3$                   & -50.8766 & -52.6578 & -54.2931\\
		${\MA}Sn_{0.25}Pb_{0.75}X_3$   & -50.8457 & -52.6239 & -54.2643\\
		${\MA}Sn_{0.5}Pb_{0.5}X_3$ [c] & -50.8197 & -52.5998 & -54.2370\\
		${\MA}Sn_{0.5}Pb_{0.5}X_3$ [d] & -50.8300 & -52.6010 & -54.2295\\
		${\MA}Sn_{0.5}Pb_{0.5}X_3$ [l] & -50.8348 & -52.6107 & -54.2553\\
		${\MA}Sn_{0.75}Pb_{0.25}X_3$   & -50.8002 & -52.5748 & -54.1986\\
		${\MA}SnX_3$                   & -50.7627 & -52.5216 & -54.1783 \\
		\hline
	\end{tabular}
	\caption{Cohesive energies $E$ per formula unit in eV for the \textit{tetragonal} phase calculated using DFT.
		The unit cell used for calculation contains 4 formula units.}
	\label{tab:Ebind_t}
\end{table}

\begin{table}
	\begin{tabular}{lccc}
		\hline
		\textbf{compound} & \textbf{X=I} & \textbf{X=Br} & \textbf{X=Cl}\\
		\hline
		${\MA}PbX_3$                   & -50.8627 & -52.7020 & -54.3528\\
		${\MA}Sn_{0.25}Pb_{0.75}X_3$   & -50.8637 & -52.6696 & -54.3129\\
		${\MA}Sn_{0.5}Pb_{0.5}X_3$ [c] & -50.8446 & -52.6416 & -54.2804\\
		${\MA}Sn_{0.5}Pb_{0.5}X_3$ [d] & -50.8398 & -52.6340 & -54.2660\\
		${\MA}Sn_{0.5}Pb_{0.5}X_3$ [l] & -50.8395 & -52.6344 & -54.2717\\
		${\MA}Sn_{0.75}Pb_{0.25}X_3$   & -50.8215 & -52.6066 & -54.2423\\
		${\MA}SnX_3$                   & -50.7913 & -52.5742 & -54.2117 \\
		\hline
	\end{tabular}
	\caption{Cohesive energies $E$ per formula unit in eV for the \textit{cubic} phase calculated using DFT.
		The unit cell used for calculation contains 8 formula units.}
	\label{tab:Ebind_c}
\end{table}

The cohesive energies of all 63 investigated structures obtained from self-consistent DFT calculations are presented in Tables \ref{tab:Ebind_o}-\ref{tab:Ebind_c}.
Since the unit cells of compounds in different phases have different symmetry and contain different number of $BX_6$ octahedra, the total binding energy of each unit cell was normalized to the energy value per formula unit, which contains one $BX_6$ octahedron and one ${\MA}$ cation.
We now analyze the energy values with respect to three parameters: the substituted halogen, the crystal symmetry of each compound, and the Sn:Pb ratio.

Comparing structures with different halogens, we can see that the cohesive energies of iodide containing structures lie in the range of $\unit[-50.90]{eV}$ and $\unit[-50.76]{eV}$, whereas the bromide structures lie in the range of $\unit[-52.72]{eV}$ and $\unit[-52.52]{eV}$, and the chloride structures lie in the range of $\unit[-54.35]{eV}$ and $\unit[-54.18]{eV}$.

There is no overlap between the energy ranges of structures with different halogens, and the energy shifts within each halogen group are relatively small ($\sim\unit[0.2]{eV}$) in comparison to those of the structures with different halogens ($\sim\unit[1.5]{eV}$).
Thus, the substitution of a halogen has the biggest influence on the cohesive energy of a structure, and the stability increases with increasing electronegativity of the halogen. 

The low-temperature orthorhombic phase of all compounds has the lowest cohesive energy, as expected.
However, for all bromide and chloride containing components, not the cubic, but the tetragonal phase with the experimentally obtained $c/a$-ratio\cite{Stoumpos2013b} has the highest cohesive energy.
Since in our systematic investigation we have used a defined set of phases that have yet been experimentally confirmed only for the ${\MA}PbI_3$ compound, this result questions the stability of the bromide and chloride containing structures in this particular tetragonal phase.
A possible explanation for this behavior could be octahedra deformations caused by the mechanism discussed in section \ref{sec:octadeform}.
For instance, the strong dependence of the crystal symmetry on the substituted halogen can be shown by the example of non-mixed ${\MA}SnCl_3$, which exhibits a different phase transition process according to PXRD experiments.\cite{Baikie2013}

With the variation of the \textit{Sn:Pb} ratio, the cohesive energy for each halogen group increases upon the gradual substitution of \textit{Pb} with \textit{Sn}, i.e., the non-mixed lead-containing compounds are the most stable and the tin-containing compounds the least stable among the \textit{Sn:Pb} mixtures.
It is worth noticing that the increase of the energy value is not linear with respect to $n_{Pb}$, which implies a regular solution with $\varepsilon \neq 0$.

One could very naively estimate the cohesive energy of a system based on the number of components of each type as
\begin{equation}\label{eq:Emu}
E^0 = N_{Pb} \, \mu_{Pb} + N_{Sn} \, \mu_{Sn}
\end{equation}
with the chemical potentials $\mu_{Pb}$ and $\mu_{Sn}$.
This formula, however, does not differentiate between different arrangements of the same $Sn:Pb$ ratio, because such a simplified model only considers the numbers of atoms $N_{Pb}$ and $N_{Sn}$.
For example, the chain, layer and diagonal arrangements all have 1:1 ratio and would therefore, according to equation (\ref{eq:Emu}), yield the same cohesive energy.
The Regular Solution Model (RSM, section \ref{sec:rsm}) based on equation (\ref{eq:Ebind}), however, explicitly accounts for the number of bonds of each type.
Therefore, it allows us to make predictions for different arrangements of $BX_6$ octahedra.
To calculate the binding energies $\mathcal{E}_{PbPb}$, $\mathcal{E}_{SnSn}$ and $\mathcal{E}_{PbSn}$ in formula (\ref{eq:Ebind}), we used the non-mixed ${\MA}PbX_3$, ${\MA}SnX_3$, and the diagonal ${\MA}Sn_{0.5}Pb_{0.5}X_3$ structures, respectively, since all of them contain only one type of bonds (see Table \ref{tab:Ncoord}).
Using equation (\ref{eq:epsilon}), these three binding energies are condensed into the single parameter, $\varepsilon$.
The power of the RSM now lies in the ability to extrapolate to the cohesive energies of structures that were not used to calculate $\varepsilon$.
This allows us, for instance, to estimate the cohesive energies of the chain and layer arrangements, as well as the ${\MA}Sn_{0.25}Pb_{0.75}X_3$ and ${\MA}Sn_{0.75}Pb_{0.25}X_3$ structures for each phase and each halogen.
We will qualitatively and quantitatively compare these results to the cohesive energies directly obtained from DFT calculations and show that, \textit{a posteriori}, the RSM provides a very good description.

In the orthorhombic phase, equation (\ref{eq:Ebind}) predicts that the layered structure is the most stable for $\varepsilon > 0$ (bromide and chloride) and the diagonal structure is more stable than chain and layer for $\varepsilon < 0$ (iodide).
This agrees with all the cohesive energies obtained from DFT calculations listed in Table \ref{tab:Ebind_o}.
In the tetragonal phase, the chain structure is, according to equation (\ref{eq:Ebind}), more stable than the diagonal reference structure in the case of $\varepsilon > 0$ (chloride), and less stable for $\varepsilon < 0$ (bromide and iodide), again in agreement with our DFT calculations (listed in Table \ref{tab:Ebind_t}).
In the cubic phase, equation (\ref{eq:Ebind}) predicts that the layered structure has a lower cohesive energy then the diagonal reference structure for $\varepsilon > 0$ (chloride and bromide) and a higher cohesive energy for $\varepsilon < 0$ (iodide), in agreement with our DFT calculations (Table \ref{tab:Ebind_c}), too.

\begin{figure}
	\includegraphics[width=0.98\linewidth]{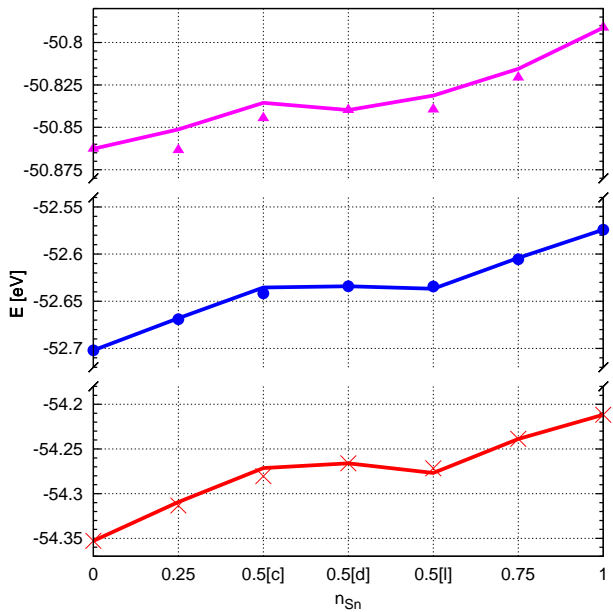}
	\caption{Comparison of the cohesive energies $E$ per formula unit  between the Regular Solution Model (solid lines) and DFT results (symbols) for the cubic phase of  iodide ${\MA}(Pb:Sn)I_3$ (top row, $\blacktriangle$), bromide ${\MA}(Pb:Sn)Br_3$ (middle row, $\bullet$) and chloride ${\MA}(Pb:Sn)Cl_3$ (bottom row, $\times$).}
	\label{fig:DFTvsRSM}
\end{figure}

We will now focus on the high-temperature cubic phase, which we consider to be most relevant for practical applications.
In Figure \ref{fig:DFTvsRSM}, we plot the obtained values from both DFT and RSM.
The root-mean-square deviation between them is only 1.4 meV and 2.2 meV for the bromide and chloride containing structures, respectively.
The deviation is higher for the iodide compounds, but since all DFT data points lie on or below the RSM curve in this case, the mixed structures are even more energetically favorable than predicted by the RSM with $\varepsilon = \unit[-2.9]{meV} < 0$.
%Hence, mixing is preferred in this case for all temperatures.

\subsection{Thermodynamic results}

\begin{table*}
	\begin{tabular}{lccc}
		\hline
		\textbf{compound} & \textbf{orthorhombic}            & \textbf{tetragonal}              & \textbf{cubic} \\
		\hline
		${\MA}(SnPb)I_3$  & $\varepsilon = \unit[-6.2]{meV}$ & $\varepsilon = \unit[-3.4]{meV}$ & $\varepsilon = \unit[-2.9]{meV}$\\\
		& -                                & -                                & - \\
		\hline
		${\MA}(SnPb)Br_3$ & $\varepsilon = \unit[8.3]{meV}$  & $\varepsilon = \unit[-3.8]{meV}$ & $\varepsilon = \unit[1.6]{meV}$\\
		& $T_{UCS} = \unit[288]{K}$            & -                                & $T_{UCS} = \unit[54]{K}$ \\
		\hline
		${\MA}(SnPb)Cl_3$ & $\varepsilon = \unit[12.2]{meV}$ & $\varepsilon = \unit[2.1]{meV}$  & $\varepsilon = \unit[5.4]{meV}$ \\
		& $T_{UCS} = \unit[425]{K}$            & $T_{UCS} = \unit[72]{K}$             & $T_{UCS} = \unit[187]{K}$ \\
		\hline
	\end{tabular}
	\caption{Mixing parameter $\varepsilon$, see equation (\ref{eq:epsilon}), and estimated upper critical solution temperatures $T_{UCS}$ of the mixed \textit{Pb:Sn} hybrid perovskites.
		Above $T_{UCS}$, mixing is favorable due to the entropy contribution.
		A dash (-) indicates that mixing is preferred at any temperature.}
	\label{tab:epsilon}
\end{table*}

The above mentioned parameter $\varepsilon$ also plays an important role in estimating the critical mixing temperatures $T_{UCS}$, which were obtained for each particular phase according to the formula (\ref{eq:t_c}).
They are summarized in Table \ref{tab:epsilon}.
To estimate whether the mixed structures containing a particular halogen actually exist, it is necessary to compare the critical mixing temperatures $T_{UCS}$ with the structural phase transition temperatures.
Mixing will only occur if the critical temperature $T_{UCS}$ is lower than the transition temperature of that phase.
The transition temperatures are known from experiments\cite{Poglitsch1987,Knop1990a,Mashiyama1998} for non-mixed ${\MA}PbX_3$ for the orthorhombic to tetragonal (tetragonal to cubic) phase transition:
they are 162K (328K) for ${\MA}PbI_3$, 146K (236K) for ${\MA}PbBr_3$ and 172K (178K) for ${\MA}PbCl_3$.
The different experiments agree within $\pm \unit[2]{K}$ and the numbers given above are averaged over the values in the cited references.
For ${\MA}SnI_3$, phase transitions are found at 111K (275K)\cite{Takahashi2011c}.
${\MA}SnBr_3$ is known to be in cubic phase at room temperature\cite{Weber1978} and calorimetric measurements\cite{Ono1991} show a transition at 229K to an experimentally unidentified (possibly tetragonal) phase.
Experiments done by Baikie \textit{et al.}\cite{Baikie2013} show that the non-mixed ${\MA}SnCl_3$ is in the cubic phase above $\unit[463]{K}$.
Below that temperature, several phases with different Bravais lattices (tricline, monocline and trigonal) are observed.
In our calculations, we indeed find a triclinic structure after a full relaxation, however, with only tiny deviations (\textless1$^\circ$) from the orthorhombic structure.

For the iodide group, $\varepsilon$ has a negative value in all phases, which means that the mixing will occur even at low temperatures.

In the case of the bromide-containing structures, only the tetragonal phases mix at all temperatures.
Bromide containing structures in the cubic phase form properly mixed arrangements only above a certain temperature, estimated by us at $T_{UCS} = \unit[54]{K}$.
Since $T_{UCS}$ is noticeably below the corresponding phase transition temperatures of 236K and 229K for non-mixed lead and tin compounds, respectively, we conclude that the bromide-containing structure in the cubic phase will always mix.
We find that mixing of bromides in the orthorhombic phase is not possible, because above the critical temperature of $T_{UCS} = \unit[288]{K}$, the orthorhombic phase is no longer stable.

We estimate the critical temperature of the orthorhombic chloride-containing structures to be $T_{UCS} = \unit[425]{K}$.
Since experiments confirmed that the orthorhombic phase is no longer stable at that temperature for the non-mixed chloride compounds, we predict that stochastic mixing in the low temperature phase will not be observed.
In the tetragonal and the cubic phase, on the other hand, the mixed structures can be stable, because the critical temperatures of both phases lie below the corresponding phase transition temperatures.

\section{Summary}
We summarize our main results based on \textit{ab-initio} DFT calculations and extrapolations to non-zero temperatures.% using the Regular Solution Model.

%Geometry
We found that geometric properties of the investigated structures are influenced by two main factors: 
on the one side chemical composition and on the other side temperature and temperature-driven phase transitions.
The unit cell volume increases with temperature, which is linked to the mobility of the organic cation.
The deformations of $BX_6$ octahedra increase with increasing electro-negativity of the halogen and are stronger for $Sn$ than for $Pb$.
The crystallographic parameters of the mixed structures fall into the range spanned by the limiting cases of the non-mixed Pb and Sn structures.
The shapes of individual $BX_6$ octahedra slightly change in mixed arrangements in order to preserve the corner-connected structure in the presence of metal ions of different size.

%Energies
The energy shifts within each halogen group are relatively small in comparison to those of the compounds with different halogens.
In particular, there is no overlap of the  cohesive-energy ranges of compounds with different halogens.
Thus, the substitution of a halogen has the biggest influence on the cohesive energy of a compound, and the stability increases with increasing electronegativity of the halogen.

%Stability
With help of the Regular Solution Model (RSM), mixing parameter $\varepsilon$ and upper critical solution temperatures $T_{UCS}$ were calculated, which then allowed to estimate the conditions at which the compounds would mix.
We predict that it will be possible to create ${\MA}(Pb:Sn)I_3$ mixtures in all three structural phases.
Our results also imply that mixing is generally not energetically preferred for the bromide and chloride containing structures, but becomes possible due to the entropy contribution above the calculated upper critical solution temperatures $T_{UCS}$ for tetragonal and cubic phases.
For the low-temperature (orthorhombic) phase of bromide and chloride containing structures, local clusters are more likely to form (if their formation is not kinetically hindered).
%Our results imply that mixing is unlikely for the low-temperature (orthorhombic) phase of bromide and chloride structures, where instead local clusters are more likely to form (if their formation is not kinetically hindered).
Thus, we conclude that for temperatures at which photovoltaic solar cells usually operate (room temperature and above), it would indeed be possible to synthesize stable mixed compounds with different \textit{Pb:Sn} stoichiometries for all three halogen groups.
%We predict that in the high-temperature cubic phase, $Pb$ and $Sn$ compounds will mix for both ${\MA}(Pb:Sn)Br_3$ and ${\MA}(Pb:Sn)Cl_3$.

\section{Acknowledgement}
We gratefully acknowledge financial support from the Th\"{u}ringer Landesgraduiertenschule PhotoGrad and the German Academic Exchange Service (DAAD).
We thank Henning Schwanbeck from the computing center of the Technische Universit\"{a}t Ilmenau for quick and professional IT support.

\bibliography{library2A}

\begin{thebibliography}{10}
\expandafter\ifx\csname urlstyle\endcsname\relax
  \providecommand{\doi}[1]{doi:\discretionary{}{}{}#1}\else
  \providecommand{\doi}{doi:\discretionary{}{}{}\begingroup
  \urlstyle{rm}\Url}\fi

\bibitem{Lee2012}
M.~M. Lee, J.~Teuscher, T.~Miyasaka, T.~N. Murakami and H.~J. Snaith;
  \emph{{Efficient hybrid solar cells based on meso-superstructured organometal
  halide perovskites.}}; Science; \textbf{338}(6107):643--647 (2012); ISSN
  1095-9203; \doi{10.1126/science.1228604}.

\bibitem{Kim2012}
H.-S. Kim, C.-R. Lee, J.-H. Im, K.-B. Lee, T.~Moehl, A.~Marchioro, S.-J. Moon,
  R.~Humphry-Baker, J.-H. Yum, J.~E. Moser, M.~Gr\"{a}tzel and N.-G. Park;
  \emph{{Lead iodide perovskite sensitized all-solid-state submicron thin film
  mesoscopic solar cell with efficiency exceeding 9\%.}}; Sci. Rep.;
  \textbf{2}:591 (2012); ISSN 2045-2322; \doi{10.1038/srep00591}.

\bibitem{Noh2013}
J.~H. Noh, S.~H. Im, J.~H. Heo, T.~N. Mandal and S.~I. Seok; \emph{{Chemical
  Management for Colorful, Efficient, and Stable Inorganic - Organic Hybrid
  Nanostructured Solar Cells}}; Nano Lett.; \textbf{13}(4):1764--1769 (2013).

\bibitem{DeBastiani2014}
M.~{De Bastiani}, V.~D'Innocenzo, S.~D. Stranks, H.~J. Snaith and A.~Petrozza;
  \emph{{Role of the crystallization substrate on the photoluminescence
  properties of organo-lead mixed halides perovskites}}; APL Mat.;
  \textbf{2}(8):081509 (2014); ISSN 2166-532X; \doi{10.1063/1.4889845}.

\bibitem{Baikie2013}
T.~Baikie, Y.~Fang, J.~M. Kadro, M.~Schreyer, F.~Wei, S.~G. Mhaisalkar,
  M.~Gratzel and T.~J. White; \emph{{Synthesis and crystal chemistry of the
  hybrid perovskite $(CH_3NH_3)PbI_3$ for solid-state sensitised solar cell
  applications}}; J. Mater. Chem. A; \textbf{1}(18):5628 (2013); ISSN
  2050-7488; \doi{10.1039/c3ta10518k}.

\bibitem{Giorgi2013}
G.~Giorgi, J.~I. Fujisawa, H.~Segawa and K.~Yamashita; \emph{{Small
  photocarrier effective masses featuring ambipolar transport in methylammonium
  lead iodide perovskite: A density functional analysis}}; J. Phys. Chem.
  Lett.; \textbf{4}(24):4213--4216 (2013); ISSN 19487185;
  \doi{10.1021/jz4023865}.

\bibitem{Weller2015}
M.~T. Weller, O.~J. Weber, P.~F. Henry, M.~D. Pumpo and T.~C. Hansen;
  \emph{{Complete structure and cation orientation in the perovskite
  photovoltaic methylammonium lead iodide between 100 and 352 K}}; Chem. Comm.;
  \textbf{51}:4180--4183 (2015); ISSN 1359-7345; \doi{10.1039/C4CC09944C}.

\bibitem{Im2012}
J.-H. Im, J.~Chung, S.-J. Kim and N.-G. Park; \emph{{Synthesis, structure, and
  photovoltaic property of a nanocrystalline 2H perovskite-type novel
  sensitizer $(CH_3CH_2NH_3)PbI_3$}}; Nanoscale Res. Lett.; \textbf{7}(1):353
  (2012); ISSN 1556-276X; \doi{10.1186/1556-276X-7-353}.

\bibitem{Safdari2014}
M.~Safdari, A.~Fischer, B.~Xu, L.~Kloo and J.~M. Gardner; \emph{{Structure and
  function relationships in alkylammonium lead(ii) iodide solar cells}}; J.
  Mater. Chem. A; \textbf{3}:9201--9207 (2015); ISSN 2050-7488;
  \doi{10.1039/C4TA06174H}.

\bibitem{Borriello2008}
I.~Borriello, G.~Cantele and D.~Ninno; \emph{{Ab initio investigation of hybrid
  organic-inorganic perovskites based on tin halides}}; Phys. Rev. B;
  \textbf{77}(23):1--9 (2008); ISSN 10980121; \doi{10.1103/PhysRevB.77.235214}.

\bibitem{Stoumpos2013b}
C.~C. Stoumpos, C.~D. Malliakas and M.~G. Kanatzidis; \emph{{Semiconducting tin
  and lead iodide perovskites with organic cations: Phase transitions, high
  mobilities, and near-infrared photoluminescent properties}}; Inorg. Chem.;
  \textbf{52}(15):9019--9038 (2013); ISSN 00201669; \doi{10.1021/ic401215x}.

\bibitem{Pellet2014}
N.~Pellet, P.~Gao, G.~Gregori, T.-Y. Yang, M.~K. Nazeeruddin, J.~Maier and
  M.~Gr\"{a}tzel; \emph{{Mixed-organic-cation perovskite photovoltaics for
  enhanced solar-light harvesting.}}; Angew. Chem. Int. Edit.;
  \textbf{53}(12):3151--7 (2014); ISSN 1521-3773; \doi{10.1002/anie.201309361}.

\bibitem{Giorgi2016}
G.~Giorgi, J.-I. Fujisawa, H.~Segawa and K.~Yamashita; \emph{{Organic-Inorganic
  Hybrid Lead Iodide Perovskite Featuring Zero Dipole Moment Guanidinium
  Cations: A Theoretical Analysis}}; J. Phys. Chem. C;
  \textbf{119}(9):4694--4701 (2015); ISSN 1932-7447;
  \doi{10.1021/acs.jpcc.5b00051}.

\bibitem{Marco2016}
N.~D. Marco, H.~Zhou, Q.~Chen, P.~Sun, Z.~Liu, L.~Meng, E.-P. Yao, Y.~Liu,
  A.~Schiffer and Y.~Yang; \emph{{Guanidinium: A Route to Enhanced Carrier
  Lifetime and Open-Circuit Voltage in Hybrid Perovskite Solar Cells}}; Nano
  Lett.; \textbf{16}(2):1009--1016 (2016); ISSN 1530-6992;
  \doi{10.1021/acs.nanolett.5b04060}.

\bibitem{Ryu2014}
S.~Ryu, J.~H. Noh, N.~J. Jeon, Y.~{Chan Kim}, W.~S. Yang, J.~Seo and S.~I.
  Seok; \emph{{Voltage output of efficient perovskite solar cells with high
  open-circuit voltage and fill factor}}; Energy. Environ. Sci.;
  \textbf{7}(8):2614 (2014); ISSN 1754-5692; \doi{10.1039/C4EE00762J}.

\bibitem{Jeon2014}
N.~J. Jeon, H.~G. Lee, Y.~C. Kim, J.~Seo, J.~H. Noh, J.~Lee and S.~I. Seok;
  \emph{{o-Methoxy substituents in spiro-OMeTAD for efficient inorganic-organic
  hybrid perovskite solar cells.}}; J. Am. Chem. Soc.;
  \textbf{136}(22):7837--7840 (2014); ISSN 1520-5126; \doi{10.1021/ja502824c}.

\bibitem{Lee2014}
J.-W. Lee, D.-J. Seol, A.-N. Cho and N.-G. Park; \emph{High-efficiency
  perovskite solar cells based on the black polymorph of {$HC(NH_2)_2PbI_3$}};
  Adv. Mater.; \textbf{26}(29):4991--4998 (2014); ISSN 1521-4095;
  \doi{10.1002/adma.201401137}.

\bibitem{Bi2016}
D.~Bi, P.~Gao, R.~Scopelliti, E.~Oveisi, J.~Luo, M.~Gr{\"{a}}tzel, A.~Hagfeldt
  and M.~K. Nazeeruddin; \emph{{High-Performance Perovskite Solar Cells with
  Enhanced Environmental Stability Based on Amphiphile-Modified
  $CH_3NH_3PbI_3$}}; Adv. Mater.; \textbf{28}(15):2910--2915 (2016); ISSN
  09359648; \doi{10.1002/adma.201505255}.

\bibitem{Yang2015}
W.~S. Yang, J.~H. Noh, N.~J. Jeon, Y.~C. Kim, S.~Ryu, J.~Seo and S.~I. Seok;
  \emph{{High-performance photovoltaic perovskite layers fabricated through
  intramolecular exchange}}; Science; \textbf{348}(6240):1234--1237 (2015);
  ISSN 0036-8075; \doi{10.1126/science.aaa9272}.

\bibitem{Mosconi2013}
E.~Mosconi, A.~Amat, M.~K. Nazeeruddin, M.~Gr\"{a}tzel and F.~{De Angelis};
  \emph{{First-principles modeling of mixed halide organometal perovskites for
  photovoltaic applications}}; J. Phys. Chem. C; \textbf{117}(27):13902--13913
  (2013); ISSN 19327447; \doi{10.1021/jp4048659}.

\bibitem{Hoke2014}
E.~T. Hoke, D.~J. Slotcavage, E.~R. Dohner, A.~R. Bowring, H.~I. Karunadasa and
  M.~D. McGehee; \emph{{Reversible photo-induced trap formation in mixed-halide
  hybrid perovskites for photovoltaics}}; Chem. Sci.; \textbf{6}(1):613--617
  (2014); ISSN 2041-6520; \doi{10.1039/C4SC03141E}.

\bibitem{Zheng2015}
F.~Zheng, H.~Takenaka, F.~Wang, N.~Z. Koocher and A.~M. Rappe;
  \emph{{First-Principles Calculation of the Bulk Photovoltaic Effect in
  $CH_3NH_3PbI_3$ and $CH_3NH_3PbI_{3-x}Cl_x$}}; J. Phys. Chem. Lett.;
  \textbf{6}(1):31--37 (2015); ISSN 1948-7185; \doi{10.1021/jz502109e}.

\bibitem{Liu2015}
J.~Liu, Y.~Shirai, X.~Yang, Y.~Yue, W.~Chen, Y.~Wu, A.~Islam and L.~Han;
  \emph{{High-Quality Mixed-Organic-Cation Perovskites from a Phase-Pure
  Non-stoichiometric Intermediate $(FAI)_{1-x}-PbI_2$ for Solar Cells}}; Adv.
  Mater.; \textbf{33}(27):4918--4923 (2015); ISSN 1521-4095;
  \doi{10.1002/adma.201501489}.

\bibitem{Seo2016}
J.~Seo, J.~H. Noh and S.~I. Seok; \emph{{Rational Strategies for Efficient
  Perovskite Solar Cells}}; Accounts Chem. Res.; \textbf{49}(3):562--572
  (2016); ISSN 1520-4898; \doi{10.1021/acs.accounts.5b00444}.

\bibitem{Wang2014b}
Y.~Wang, T.~Gould, J.~F. Dobson, H.~Zhang, H.~Yang, X.~Yao and H.~Zhao;
  \emph{{Density functional theory analysis of structural and electronic
  properties of orthorhombic perovskite $CH_3NH_3PbI_3$}}; Phys. Chem. Chem.
  Phys.; \textbf{16}(4):1424--1429 (2014); ISSN 1463-9084;
  \doi{10.1039/c3cp54479f}.

\bibitem{Hao2014b}
F.~Hao, C.~C. Stoumpos, D.~H. Cao, R.~P.~H. Chang and M.~G. Kanatzidis;
  \emph{{Lead-free solid-state organic–inorganic halide perovskite solar
  cells}}; Nat. Photonics; \textbf{8}(6):489--494 (2014); ISSN 1749-4885;
  \doi{10.1038/nphoton.2014.82}.

\bibitem{Noel2014}
N.~K. Noel, S.~D. Stranks, A.~Abate, C.~Wehrenfennig, S.~Guarnera, A.-A.
  Haghighirad, A.~Sadhanala, G.~E. Eperon, S.~K. Pathak, M.~B. Johnston,
  A.~Petrozza, L.~M. Herz and H.~J. Snaith; \emph{{Lead-Free Organic-Inorganic
  Tin Halide Perovskites for Photovoltaic Applications}}; Energy Environ. Sci.;
  \textbf{7}:3061--3068 (2014); ISSN 1754-5692; \doi{10.1039/C4EE01076K}.

\bibitem{Bernal2014}
C.~Bernal and K.~Yang; \emph{{First-Principles Hybrid Functional Study of the
  Organic-Inorganic Perovskites $CH_3NH_3SnBr_3$ and $CH_3NH_3SnI_3$}}; J.
  Phys. Chem. C; \textbf{118}(42):24383--24388 (2014).

\bibitem{Hao2014}
F.~Hao, C.~C. Stoumpos, R.~P.~H. Chang and M.~G. Kanatzidis; \emph{{Anomalous
  band gap behavior in mixed Sn and Pb perovskites enables broadening of
  absorption spectrum in solar cells}}; J. Am. Chem. Soc.;
  \textbf{136}(22):8094--8099 (2014); \doi{10.1021/ja5033259}.

\bibitem{Zuo2014}
F.~Zuo, S.~T. Williams, P.~W. Liang, C.~C. Chueh, C.~Y. Liao and A.~K.~Y. Jen;
  \emph{{Binary-metal perovskites toward high-performance planar-heterojunction
  hybrid solar cells}}; Adv. Mater.; \textbf{26}(37):6454--6460 (2014); ISSN
  09359648; \doi{10.1002/adma.201401641}.

\bibitem{Ogomi2014}
Y.~Ogomi, A.~Morita, S.~Tsukamoto, T.~Saitho, N.~Fujikawa, Q.~Shen, T.~Toyoda,
  K.~Yoshino, S.~S. Pandey and S.~Hayase; \emph{{$CH_3NH_3Sn_{x}Pb_{x-1}I_3$
  Perovskite Solar Cells Covering up to 1060 nm}}; J. Phys. Chem. Lett.;
  \textbf{5}(6):1004--1011 (2014).

\bibitem{Mosconi2014}
E.~Mosconi, P.~Umari and F.~{De Angelis}; \emph{{Electronic and optical
  properties of mixed Sn–Pb organohalide perovskites: a first principles
  investigation}}; J. Mater. Chem. A; \textbf{3}:9208--9215 (2015); ISSN
  2050-7488; \doi{10.1039/C4TA06230B}.

\bibitem{Kresse1993}
G.~Kresse and J.~Hafner; \emph{{Ab initio molecular dynamics for liquid
  metals}}; Phys. Rev. B; \textbf{47}(1):558--561 (1993);
  \doi{10.1103/PhysRevB.47.558}.

\bibitem{Kresse1996}
G.~Kresse and J.~Furthm\"{u}ller; \emph{{Efficient iterative schemes for ab
  initio total-energy calculations using a plane-wave basis set}}; Phys. Rev.
  B; \textbf{54}(16):11169--11186 (1996); \doi{10.1103/PhysRevB.54.11169}.

\bibitem{Bloechl1994}
P.~E. Bloechl; \emph{{Projector augmented-wave method}}; Phys. Rev. B;
  \textbf{50}(24):17953--17979 (1994); \doi{10.1103/PhysRevB.50.17953}.

\bibitem{Kresse1999}
G.~Kresse and D.~Joubert; \emph{{From ultrasoft pseudopotentials to the
  projector augmented-wave method}}; Phys. Rev. B; \textbf{59}(3):1758--1775
  (1999); \doi{10.1103/PhysRevB.59.1758}.

\bibitem{Perdew1996}
J.~P. Perdew, K.~Burke and M.~Ernzerhof; \emph{{Generalized Gradient
  Approximation Made Simple}}; Phys. Rev. Lett.; \textbf{77}(18):3865--3868
  (1996); ISSN 0031-9007; \doi{10.1103/PhysRevLett.77.3865}.

\bibitem{Haasen1996}
P.~Haasen; \emph{{Physical Metallurgy}}; Cambridge University Press; 3 edn.
  (1996); ISBN 0521550920.

\bibitem{Flory1951}
P.~J. Flory and W.~R. Krigbaum; \emph{{Thermodynamics of High Polymer
  Solutions}}; Annu. Rev. Phys. Chem.; \textbf{2}(1):383--402 (1951); ISSN
  0066-426X; \doi{10.1146/annurev.pc.02.100151.002123}.

\bibitem{Huggins1943}
M.~L. Huggins; \emph{{Thermodynamic properties of solutions of high polymers:
  The empirical constant in the activity equation}}; Ann. N. Y. Acad. Sci.;
  \textbf{44}(4):431--443 (1943); ISSN 00778923;
  \doi{10.1111/j.1749-6632.1943.tb52763.x}.

\bibitem{Tricker1978}
M.~Tricker and J.~Donaldson; \emph{{Comments on the structure, bonding and
  $^{119}Sn$ M\"{o}ssbauer parameters of tin(II) derivatives of the type
  $MSnX_3$}}; Inorg. Chim. Acta.; \textbf{31}(C):L445--L446 (1978); ISSN
  00201693; \doi{10.1016/S0020-1693(00)94957-0}.

\bibitem{Poglitsch1987}
A.~Poglitsch and D.~Weber; \emph{{Dynamic disorder in
  methylammoniumtrihalogenoplumbates (II) observed by millimeter-wave
  spectroscopy}}; J. Chem. Phys; \textbf{87}(11):6373 (1987); ISSN 00219606;
  \doi{10.1063/1.453467}.

\bibitem{Frost2014}
J.~M. Frost, K.~T. Butler and A.~Walsh; \emph{{Molecular ferroelectric
  contributions to anomalous hysteresis in hybrid perovskite solar cells}}; APL
  Mat.; \textbf{2}(8):081506 (2014); ISSN 2166-532X; \doi{10.1063/1.4890246}.

\bibitem{Murnaghan1944}
F.~D. Murnaghan; \emph{{The Compressibility of Media under Extreme Pressures}};
  Proc. Natl. Acad. Sci. USA; \textbf{30}(9):244--247 (1944); ISSN 0027-8424;
  \doi{10.1073/pnas.30.9.244}.

\bibitem{Woodward1997}
P.~M. Woodward; \emph{{Octahedral Tilting in Perovskites. I. Geometrical
  Considerations}}; Acta Crystallogr. B; \textbf{53}(1):32--43 (1997); ISSN
  01087681; \doi{10.1107/S0108768196010713}.

\bibitem{Garcia-Fernandez2010}
P.~Garcia-Fernandez, J.~Aramburu, M.~Barriuso and M.~Moreno; \emph{{Key Role of
  Covalent Bonding in Octahedral Tilting in Perovskites}}; J. Phys. Chem.
  Lett.; \textbf{1}(3):647--651 (2010); ISSN 1948-7185;
  \doi{10.1021/jz900399m}.

\bibitem{Knop1990a}
O.~Knop, R.~E. Wasylishen, M.~A. White, T.~S. Cameron and M.~J. M.~V. Oort;
  \emph{{Alkylammonium lead halides. Part 2. $CH_3NH_3PbX_3$ (X = Cl, Br, I)
  perovskites: cuboctahedral halide cages with isotropic cation
  reorientation}}; Can. J. Chem.; \textbf{68}(3):412--422 (1990); ISSN
  0008-4042; \doi{10.1139/v90-063}.

\bibitem{Mashiyama1998}
H.~Mashiyama; \emph{{Disordered Cubic Perovskite Structure of $CH_3NH_3PbX_3$
  (X = Cl, Br, I)}}; J. Korean Phys. Soc.; \textbf{32}:S156--S158 (1998); ISSN
  03744884.

\bibitem{Takahashi2011c}
Y.~Takahashi, R.~Obara, Z.-Z. Lin, Y.~Takahashi, T.~Naito, T.~Inabe,
  S.~Ishibashi and K.~Terakura; \emph{{Charge-transport in tin-iodide
  perovskite $CH_3NH_3SnI_3$: origin of high conductivity}}; Dalton Trans.;
  \textbf{40}(20):5563--5568 (2011); ISSN 1477-9226; \doi{10.1039/c0dt01601b}.

\bibitem{Weber1978}
D.~Weber; \emph{{$CH_3NH_3SnBr_xI_{3-x}$ (x=0-3), ein Sn(II)-System mit
  kubischer Perowskitstruktur}}; Z. Naturforsch. B; \textbf{33}(8):862--865
  (1978).

\bibitem{Ono1991}
N.~Onoda-Yamamuro, T.~Matsuo and H.~Suga; \emph{{Thermal, electric, and
  dielectric properties of $CH_3NH_3SnBr_3$ at low temperatures}}; J. Chem.
  Thermodyn.; \textbf{23}(39):987--999 (1991).

\end{thebibliography}
\bibliographystyle{myarxiv}%unsrt

\end{document}